\documentclass[twocolumn]{aastex631}

\usepackage{CJKutf8}
\usepackage{amsmath}
\usepackage{enumitem}
\usepackage{booktabs,pbox}
\usepackage{bm}
\usepackage{comment}
\usepackage[caption=false]{subfig}

\newcommand\PMO{Purple Mountain Observatory, Chinese Academy of Sciences, Nanjing 210023, China}
\newcommand\USTC{School of Astronomy and Space Sciences, University of Science and Technology of China, Hefei 230026, China}
\newcommand\NJU{School of Astronomy and Space Science, Nanjing University, Nanjing 210023, China}
\newcommand\NJULab{Key Laboratory of Modern Astronomy and Astrophysics (Nanjing University), Ministry of Education, Nanjing 210023, China}
\newcommand{\ICRA}{ICRANet, Piazza della Repubblica 10, I-65122 Pescara, Italy}

\begin{document}

\title{Probing Thermal Electrons in GRB Afterglows}

\author[0000-0002-2035-4688]{Hao-Xuan Gao}
\affiliation{\PMO}

\author[0000-0002-9583-2947]{Jin-Jun Geng}\thanks{E-mail: jjgeng@pmo.ac.cn}
\affiliation{\PMO}

\author[0000-0002-9583-2947]{Tian-Rui Sun}
\affiliation{\PMO}

\author[0000-0002-9583-2947]{Liang Li}
\affiliation{\ICRA}

\author[0000-0002-9583-2947]{Yong-Feng Huang}\thanks{E-mail: hyf@nju.edu.cn}
\affiliation{\NJU}
\affiliation{\NJULab}

\author[0000-0002-9583-2947]{Xue-Feng Wu}\thanks{E-mail: xfwu@pmo.ac.cn}
\affiliation{\PMO}
\affiliation{\USTC}

\begin{abstract}
Particle-in-cell simulations have unveiled that shock-accelerated electrons do not follow a pure power-law distribution, but have an additional low-energy ``thermal'' part, which owns a considerable portion of the total energy of electrons.
Investigating the effects of these thermal electrons on gamma-ray burst (GRB) afterglows may provide valuable insights into the particle acceleration mechanisms.
We solve the continuity equation of electrons in the energy space, from which multi-wavelength afterglows are derived by incorporating processes including synchrotron radiation, synchrotron self-absorption, synchrotron self-Compton scattering, and gamma-gamma annihilation.
First, there is an underlying positive correlation between temporal and spectral indices due to the cooling of electrons.
Moreover, thermal electrons would result in the simultaneous non-monotonic variation in both spectral and temporal indices at multi-wavelength, which could be individually recorded by the 2.5-meter Wide Field Survey Telescope and Vera Rubin Observatory Legacy Survey of Space and Time (LSST).
The thermal electrons could also be diagnosed from afterglow spectra by synergy observation in the optical (with LSST) and X-ray (with the Microchannel X-ray Telescope on board the Space Variable Objects Monitor) bands.
Finally, we use Monte Carlo simulations to obtain the distribution of peak flux ratio ($R_{\rm X}$) between soft and hard X-rays, and of the time delay ($\Delta t$) between peak times of soft X-ray and optical light curves.
The thermal electrons significantly raise the upper limits of both $R_{\rm X}$ and $\Delta t$.
Thus the distribution of GRB afterglows with thermal electrons is more scattered in the $R_{\rm X} - \Delta t$ plane.
\end{abstract}

\keywords{acceleration of particles -- gamma-ray burst: general -- radiation mechanisms: general -- relativistic processes}

\section{Introduction} \label{sec:intro}

Gamma-ray bursts (GRBs) are the most powerful stellar explosions in the Universe.
They generally lasts for several milliseconds to a few minutes, releasing a huge
amount of energy in hard X-rays and gamma-rays. GRBs are widely believed to
originate from relativistic outflows ejected during the collapse of massive stars
or the merger of binary compact stars \citep{Rees92,Paczynski98,Eichler89,Piran04,Kumar15}.

The prompt emission is expected to arise from the internal energy dissipation, which is caused by internal shocks resulting from the collision between different parts of the outflow \citep{Rees94,Paczynski94} or magnetic reconnection \citep{Spruit01,ZhangYang11}, or from the Comptonized quasi-thermal emission originating from the photosphere of the outflow \citep{Thompson94,Ghisellini99,Meszaros00,Peer08,Peer11,Lundman13,Deng14}.
Subsequently, the jet continue to interact with the circum-burst medium \citep{Meszaros93}, producing a
forward shock that propagates into the surrounding medium \citep{Meszaros97,Dai98,Sari99}, and a reverse shock that travels into the ejecta \citep{Sari99,Kobayashi00a,Kobayashi00b,YuDai07,Geng18c}.
Synchrotron radiation from the shock-accelerated electrons will then lead to a long-lasting afterglow observable in softer bands, ranging from radio to X-rays \citep{Sari98}.
Currently, many key questions regarding GRB jets still remain unsolved, including
the composition of the jets, the radiation mechanism responsible for the prompt
emission, and the distance of the emitting region from the central
engine.

The emissions of GRB afterglows are closely connected with the acceleration of electrons by relativistic shocks, which, however is still poorly understood.
In most theoretical studies of GRB afterglows, it is generally assumed that the
downstream particles carry a fraction $\epsilon_{\mathrm{e}}$ of the shock-dissipated
energy and follow a simple power-law distribution \citep{Paczynski93,Sari98,Panaitescu05,Li19,Medina23}.
However, recent first-principle particle-in-cell (PIC) simulations suggest that only a portion of
electrons may be efficiently accelerated, while the rest form a relativistic Maxwellian distribution
centered at a lower energy \citep[e.g.,][]{Park15,Crumley19,Spitkovsky08a,Spitkovsky08b,Martins09,Sironi13}. Especially, \citet{Eichler05} performed
a rudimentary treatment of synchrotron cooling and self-absorption and argued that the thermal electrons
could be identified through early ($\leq$ 10 hr) radio afterglow observations. \citet{Giannios09} delved
into the effects of the thermal electron component on afterglow spectra, light curves, and prompt emission.
They found that the thermal electrons will lead to a non-monotonic hard-soft-hard variation in the
X-ray spectral index. \citet{Pennanen14} suggested that a thermal electron injection could lead
to a flattening in the X-ray light curve which should be detectable in a wind-type environment.
Moreover, thermal electrons contribute to additional opacity due to synchrotron self-absorption,
leading to a significant increase in the synchrotron self-absorption frequency (typically by a
factor of 10 -- 100) \citep{Ressler17,Warren17,Warren18}. \citet{Margalit21} demonstrated that
the shock speed plays a pivotal role in the effect of thermal electrons. Specifically, they
stressed that for mildly relativistic shocks connected to AT2018cow-like events, thermal
electrons should play the dominant role at the peak time of the afterglow.

The thermal electron component in the downstream is closely related to the properties of the shock, and its energy proportion is markedly influenced by the shock magnetization.
Investigating the effect of thermal electrons in GRB afterglows may provide valuable clues on the particle acceleration process.
Recently, we have studied the continuity equation of electrons in the energy space by using the constrained interpolation profile method \citep{Yabe01}.
The cooling process of electrons accelerated by internal shocks is studied, and the evolution patterns of the peak energy of the prompt emission as well as the inverse Compton scattering spectra
are investigated \citep{Geng18b,Zhang19,Gao21}. Here, we will further study the effect of thermal electrons on the multi-wavelength afterglow of GRBs. For this purpose, a numerical code
is developed to deal with various factors involved in the continuity equation, such as the
synchrotron self-absorption process, the synchrotron self-Compton scattering, and the
gamma-gamma annihilation.

The structure of our paper is organized as follows. A brief description of our model
is presented in Section \ref{sec:models}. Section \ref{sec:calculations} explores the
way to identify the existence of thermal electrons in GRB afterglows.
In Section \ref{sec:correlation}, numerical results are presented by taking various
parameter sets. The observational effects of thermal electrons are further discussed
in Section \ref{sec:observation}. Finally, we summarize our study in Section \ref{sec:conclusions}.

\section{Model of the Afterglow} \label{sec:models}

An external shock will be excited when the outflow ejected by the central engine
of a GRB interacts with the circum-burst medium. Synchrotron radiation of electrons
accelerated by the shock produces the observed afterglow \citep{Waxman97,Wijers97,Sari98,Sari99,Huang99,Huang00c,Geng14,Geng16a,Gao22}.
During this process, a fraction $\epsilon_{\mathrm{e}}$ of the shock energy is transferred to
electrons, while a fraction $\epsilon_{B}$ goes to the magnetic field. As mentioned in the
Introduction, the distribution of the downstream electrons should be a superposition of two
components, i.e., a Maxwellian component and a power-law component: 
\begin{equation}
Q\left(\gamma_{\mathrm{e}}^{}, t^{}\right)=
\begin{cases}
Q_{\mathrm{inj}}(t^{}) N_{\mathrm{e}}^{\mathrm{th}}(\gamma_{\mathrm{e}}, \Theta), \quad \quad \quad\quad\quad\gamma_{\mathrm{e}} \leq \gamma_{\mathrm{nth}}, \\
Q_{\mathrm{inj}}(t^{}) N_{\mathrm{e}}^{\mathrm{th}}\left(\gamma_{\mathrm{nth}},\Theta\right){\left(\frac{\gamma_{\mathrm{e}}} {\gamma_{\mathrm{nth}}}\right)}^{-p}, \gamma_{\mathrm{e}}>\gamma_{\mathrm{nth}},
\end{cases}
\label{eq:distribution}
\end{equation}
where $Q_{\mathrm{inj}}$ is the normalized injection rate, the Maxwellian component is expressed
as $N_{\mathrm{e}}^{\mathrm{th}}(\gamma_{\mathrm{e}}, \Theta) =
\gamma_{\mathrm{e}}^2 \exp (-\gamma_{\mathrm{e}} / \Theta) / 2 \Theta^3$,
and the dimensionless temperature of thermal electrons is denoted by $\Theta\equiv\frac{k_{B}T_{e}}{m_{e}c^{2}}$.
As the jet interacts with the circumburst medium, it gradually slows down, causing both the bulk
Lorentz factor ($\Gamma$) and the magnetic field to decrease. Correspondingly, the average
Lorentz factor of electrons evolves as
\begin{equation}
\langle\gamma_{\mathrm{e}}\rangle = \frac{1}{Q_{\mathrm{inj}}}\int_1^{\infty} \gamma_{\mathrm{e}}  Q\left(\gamma_{\mathrm{e}}^{}, t^{}\right) d \gamma_{\mathrm{e}}=\epsilon_{\mathrm{e}} (\Gamma-1) \frac{m_{\mathrm{p}}}{m_{\mathrm{e}}}.
\end{equation}

The electrons lose energy through synchrotron radiation and adiabatic cooling. As a result, their
distribution also changes with time, which can be calculated by solving the continuity equation of \citep{Longair11}
\begin{equation}
\frac{\partial}{\partial t^{\prime}}\left(\frac{d N_{\mathrm{e}}}{d \gamma_{\mathrm{e}}^{\prime}}\right)+\frac{\partial}{\partial \gamma_{\mathrm{e}}^{\prime}}\left[\dot{\gamma}_{\mathrm{e}, \mathrm { tot }}^{\prime}\left(\frac{d N_{\mathrm{e}}}{d \gamma_{\mathrm{e}}^{\prime}}\right)\right]=Q\left(\gamma_{\mathrm{e}}^{\prime}, t^{\prime}\right).
\end{equation}
Then we can calculate the power of synchrotron radiation at frequency $\nu^{\prime}$ (the superscript of prime
denotes the quantities in the comoving frame hereafter) \citep{Rybicki79}:
\begin{equation}
P_{\mathrm{syn}}^{\prime}\left(\nu^{\prime}\right)=\frac{\sqrt{3} q_{\mathrm{e}}^3 B^{\prime}}{m_{\mathrm{e}} c^2} \int_{\gamma_{\mathrm{e}, \min }^{\prime}}^{\gamma_{\mathrm{e}, \max }^{\prime}}\left(\frac{d N_{\mathrm{e}}}{d \gamma_{\mathrm{e}}^{\prime}}\right) F\left(\frac{\nu^{\prime}}{\nu_{\mathrm{c}}^{\prime}}\right) d \gamma_{\mathrm{e}}^{\prime}.
\end{equation}

The radio flux will be suppressed by the synchrotron self-absorption (SSA) effect, which is included
in our calculations. We also consider the up-scatter of photons by relativistic electrons, i.e. the
synchrotron self-Compton (SSC) scattering, the power of which is \citep{Blumenthal70,Zhang19}
\begin{equation}
\begin{aligned}
P_{\mathrm{SSC}}^{\prime}\left(\nu_{\mathrm{ic}}^{\prime}\right)= & \frac{3 \sigma_{\mathrm{T}} c h \nu_{\mathrm{ic}}^{\prime}}{4} \int_{\nu_{\min }^{\prime}}^{\nu_{\max }^{\prime}} \frac{n^{\prime}\left(\nu^{\prime}\right) d \nu^{\prime}}{\nu^{\prime}} \\
& \times \int_{\gamma_{\mathrm{e}, \text { min }}^{\prime}}^{\gamma_{\mathrm{e}, \max }^{\prime}} \frac{F(q, g)}{\gamma_e^{\prime 2}} \frac{d N_{\mathrm{e}}^{\prime}\left(\gamma_{\mathrm{e}}^{\prime}\right)}{d \gamma_{\mathrm{e}}^{\prime}} d \gamma_{\mathrm{e}}^{\prime},
\end{aligned}
\end{equation}
where $F(q, g)=2 q \ln q+(1+2 q)(1-q)+\frac{(4 q g)^2}{2(1+4 q g)}(1-q)$, $\quad g=\frac{\gamma_{\mathrm{e}}^{\prime} h \nu^{\prime}}{m_{\mathrm{e}} c^2}$, $w=\frac{h \nu_{\mathrm{ic}}^{\prime}}{\gamma_{\mathrm{e}}^{\prime} m_{\mathrm{e}} c^2}$, and $q=\frac{w}{4 g(1-w)}$.
The gamma-gamma annihilation further attenuates the flux, with a power of
\begin{equation}
P_{\gamma \gamma}^{\prime}\left(\nu^{\prime}\right)=P^{\prime}\left(\nu^{\prime}\right) \frac{1-e^{-\tau_{\gamma \gamma}(\nu^{\prime})}}{\tau_{\gamma \gamma}(\nu^{\prime})}.
\end{equation}
The optical depth of gamma-gamma annihilation ($\tau_{\gamma \gamma}(\nu)$) can then be expressed as \citep{Gould67},
\begin{equation}
\tau_{\gamma \gamma}(\nu^{\prime})=\frac{1}{4 \pi R \Gamma c} \frac{2}{h^2}\left(\frac{m_e^2 c^4}{h \nu^{\prime}}\right)^2 \int_{\nu_{\min }^{\prime}}^{\nu_{\max }^{\prime}} \frac{\sigma_{\gamma \gamma}\left(\nu^{\prime}, \tilde{\nu^{\prime}}\right) P^{\prime}\left(\tilde{\nu^{\prime}}\right)}{h{\tilde{\nu^{\prime}}}^3} d \tilde{\nu^{\prime}},
\end{equation}
where $\sigma_{\mathrm{\gamma \gamma}} $ is the cross section of gamma-gamma annihilation between a photon with a frequency of $\nu^{\prime}$ and another photon with a frequency of $\tilde{\nu^{\prime}}$,
$\tilde{\nu^{\prime}}$ represents the frequency of a photon in the background radiation field of $P^{\prime}\left(\tilde{\nu^{\prime}}\right)$,
and $P^{\prime}\left(\tilde{\nu^{\prime}}\right)=P_{\mathrm{syn}}^{\prime}\left(\tilde{\nu^{\prime}}\right)+
P_{\mathrm{SSC}}^{\prime}\left(\tilde{\nu^{\prime}}\right)$.
Finally, note that to obtain the spectrum in the observer's frame, we need to sum up the emission from electrons on the equal-arrival-time surface \citep{Geng16a}.

\section{Numerical Results} \label{sec:calculations}

In this section, we calculate the afterglow of GRBs in two different cases of circum-burst
medium, i.e. the homogeneous interstellar medium and the stellar wind medium.
In the standard forward shock model, the afterglow flux is a function of both frequency
and time that could be expressed as $F_{\nu} \propto \nu^{-\beta} t^{-\alpha}$.
The values of the two power-law indices, $\alpha$ and $\beta$, are different in different
spectral regimes \citep[more details could be found in][]{Sari98,Zhang06,Gao13a,Zhang18}.
The situation is more complex when the evolution of the continuity equation is considered.
Here, we will first try to derive the relation between $\alpha$ and $\beta$ under some
simplified assumptions.

\subsection{An Analytic Equation}
\label{sec:Analytic}

We first consider the simplified physical scenario, in which the cooling rate ($\dot{\gamma_{\mathrm{e}}^{\prime}}$) of electrons in the energy space is independent of electron energy and there is no external electron injection.
The number of electrons is conserved in the cooling process, thus electrons that cross the position at $\gamma_{\mathrm{e}}^{\prime}+\dot{\gamma_{\mathrm{e}}^{\prime}}dt^{\prime}$ at moment $t^{\prime}$ will pass the position at $\gamma_{\mathrm{e}}^{\prime}$ at moment $t^{\prime}+dt^{\prime}$, i.e.,
\begin{equation}
f(\gamma_{\mathrm{e}}^{\prime},t^{\prime}) d \gamma_{\mathrm{e}}^{\prime}= f(\gamma_{\mathrm{e}}^{\prime} + \dot{\gamma_{\mathrm{e}}^{\prime}} dt^{\prime}, t^{\prime}-dt^{\prime})d \gamma_{\mathrm{e}}^{\prime},
\label{eq:simplest}
\end{equation}
where $f(\gamma_{\mathrm{e}}^{\prime},t^{\prime})$ is the electron distribution.
Equation (\ref{eq:simplest}) is equivalent to
\begin{equation}
\frac{\partial f}{\partial t^{\prime}}=\dot{\gamma_{\mathrm{e}}^{\prime}}\frac{\partial f}{\partial \gamma_{\mathrm{e}}^{\prime}},
\label{eq:property}
\end{equation}
which indicates that the time variation rate of the electron distribution in the phase space at the point of ($\gamma_{\mathrm{e}}^{\prime}$, $t^{\prime}$) is proportional to its spatial gradient.
Moreover, if we assume that electrons of a specific energy can only emit photons of a specific frequency under the same magnetic field, the property of the electron distribution shown by Equation (\ref{eq:property}) could be found in the observed afterglow spectrum in form of
\begin{equation}
\frac{\partial F_{\nu_{\mathrm{obs}}}}{\partial t_{\mathrm{obs}}} \propto \frac{\partial F_{\nu_{\mathrm{obs}}}}{\partial \nu_{\mathrm{obs}}}.
\end{equation}
Under the power-law approximation to the temporal and spectral feature of the afterglow, we can infer a relationship of $\alpha \propto \beta$.
However, the cooling rates of electrons depend on their energy and the external electron injection occurs during the cooling process.
A more complicated equation will be obtained, which is
\begin{equation}
\alpha= \frac{t_{\mathrm{obs}}}{\nu_{\mathrm{obs}}}\Lambda \beta-
\Omega t_{\mathrm{obs}}-\frac{t_{\mathrm{obs}}}{F_{\nu_{\mathrm{obs}}}}\frac{\Gamma \mathcal{D}}{4\pi D_{\mathrm{L}}^{2}}\hat{Q},
\label{eq:ab}
\end{equation}
where $\mathcal{D}=1/\Gamma(1-\beta_{\mathrm{j}})$ is the Doppler factor, $\beta_{\mathrm{j}}$ is the dimensionless velocity of the shock, and more derivation details can be found in Appendix \ref{apdx:Eqder}.

The right-hand side of Equation (\ref{eq:ab}) consists of three terms: the first term arises
from the advection of electrons in the energy space, the second term comes from the electron
convection, and the third term is a consequence of the electron injection. Equation (\ref{eq:ab})
is derived based on the assumption that a single electron emits photons exclusively at a
particular characteristic frequency through the synchrotron radiation mechanism. However, for
radiation mechanism other than the synchrotron radiation, as long as the emission is essentially
monochromatic and the electrons cool down mainly through radiation, a result similar to
Equation (\ref{eq:ab}) can still be obtained. Equation (\ref{eq:ab}) is largely independent of the
specific radiation mechanism, and the correlation between $\alpha$ and $\beta$ is predominantly
determined by the cooling process of electrons.

Equation (\ref{eq:ab}) indicates the relationship between the two indices can be expressed
as $\alpha = a \cdot \beta - b$, where both of the two coefficients, $a$ and $b$, depend
on the parameters of $\Gamma$, $B^{\prime}$, $\nu_{\mathrm{obs}}$, and $t_{\mathrm{obs}}$.
Note that $\beta$ generally is of the order of $\sim1$. When $b$ is significantly less
than $a$, the sign of $\alpha$ would be the same as that of $\beta$, and a positive
correlation exists between them. However, when $b \gg a$, the sign of $\alpha$ will be
opposite to that of $\beta$.

\begin{figure}
	\centering\includegraphics[scale=.3]{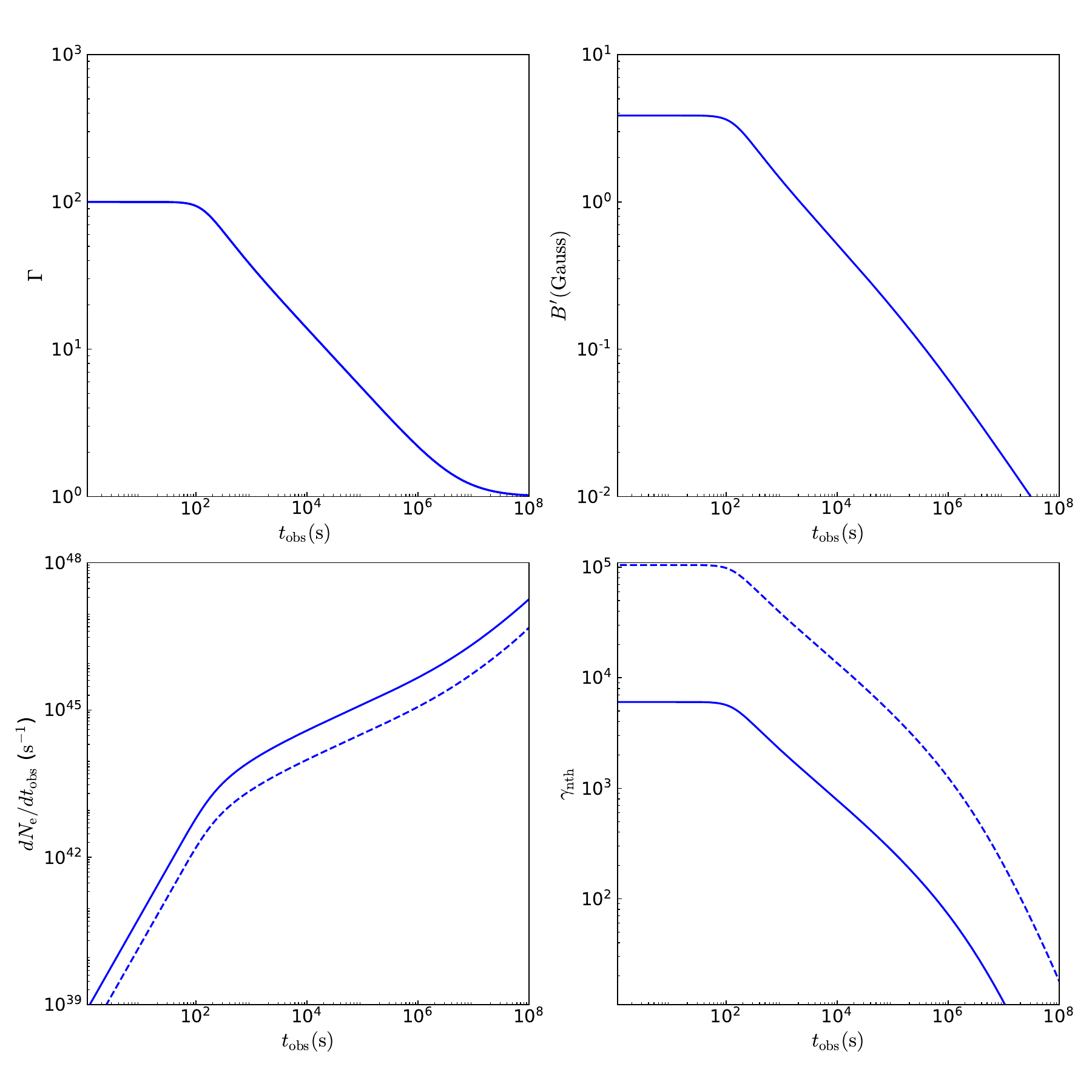}
    \caption{The evolution of the bulk Lorentz factor ($\Gamma$, upper-left panel),
    the co-moving frame magnetic field strength ($B^{\prime}$, upper-right panel),
    the injection rate of electrons ($dN_{\mathrm{e}}/dt_{\mathrm{obs}}$, lower-left
    panel), and the minimum Lorentz factor of the injected non-thermal electrons
    ($\gamma_{\mathrm{nth}}$, lower-right panel). This figure is plotted for a
    jet expanding in a homogeneous interstellar medium with $\delta=1$ (solid curve)
    or $\delta=0.2$ (dashed curve).   }
    \label{fig:dyna1}
\end{figure}

\begin{figure*}
\centerline{\includegraphics[width=0.7\textwidth,trim= 0 0 0 0 , clip]{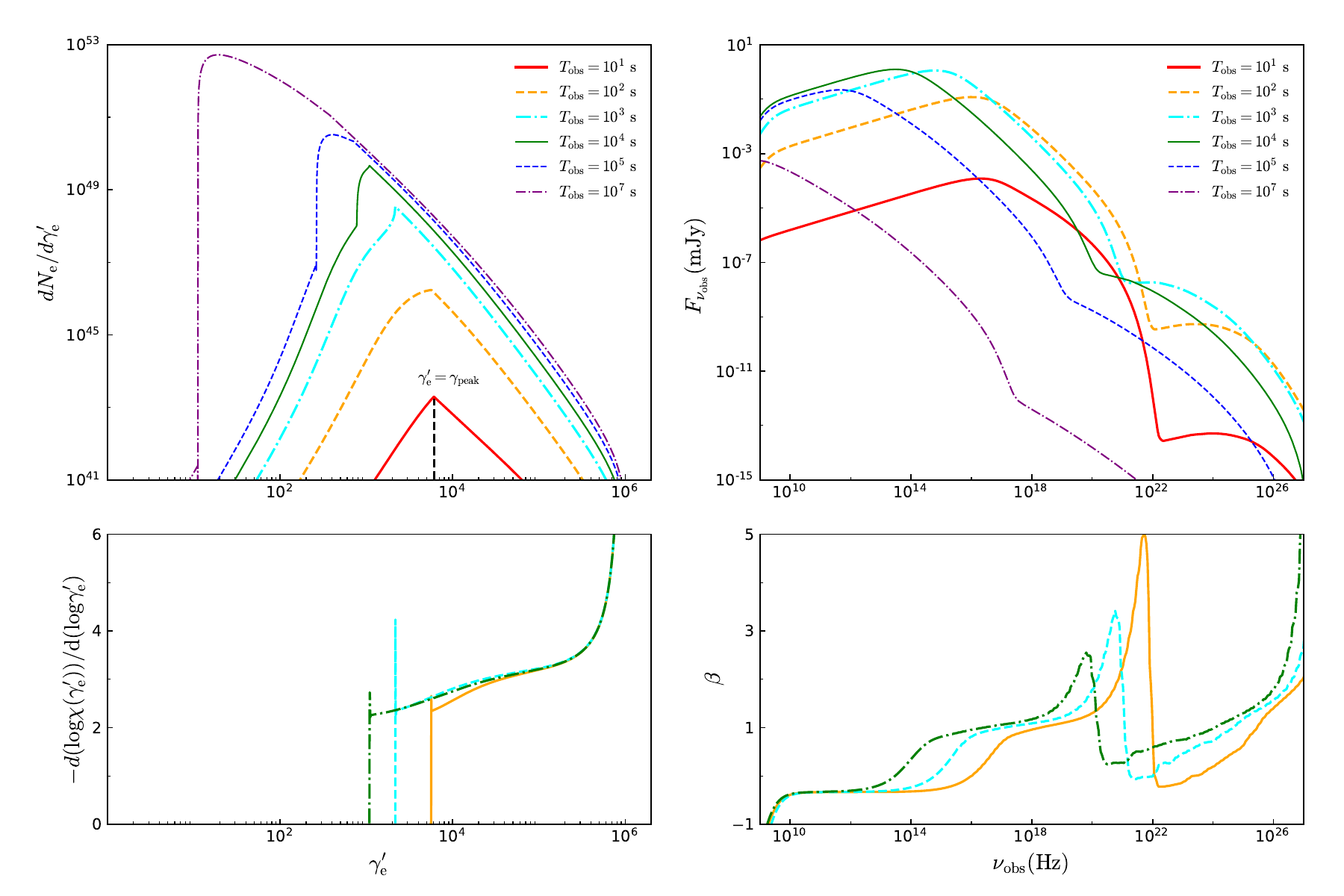}}
\caption{The evolution of the distribution function of electrons (upper-left panel) and the
afterglow spectrum (upper-right panel) in the interstellar medium case with $\delta=1$.
The power-law indices of the electron distribution function (i.e., $\chi(\gamma_{\mathrm{e}}^{\prime})=d N_{\mathrm{e}} /d \gamma_{\mathrm{e}}^{\prime}$) and the spectral indices of afterglow spectrum are shown
correspondingly in the lower panels. The characteristic Lorentz factor $\gamma_\mathrm{peak}$
is marked in the upper-left panel}.
\label{fig:ED1}
\end{figure*}

\begin{figure}
	\centering\includegraphics[scale=.4]{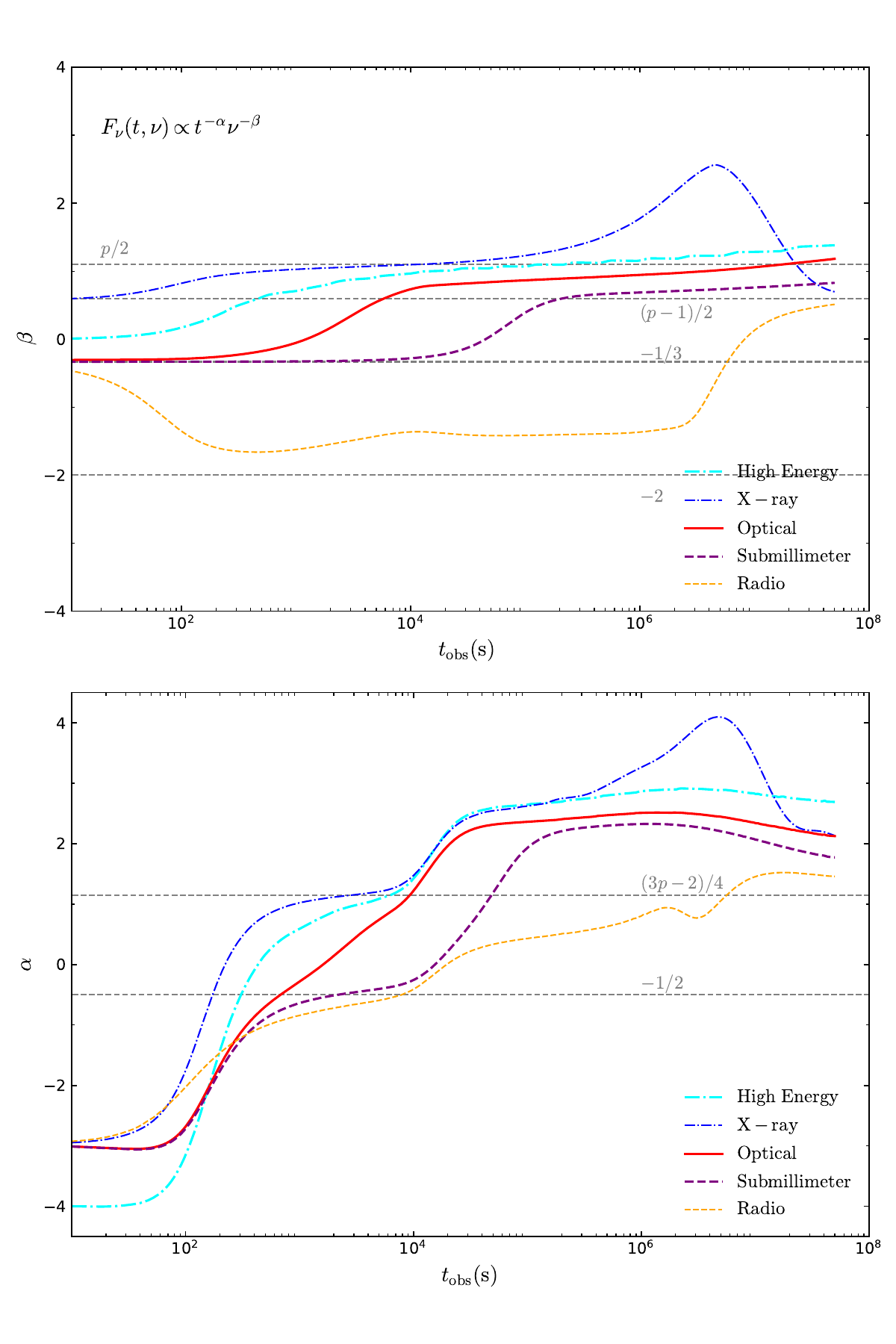}
    \caption{Evolution of the spectral index $\beta$ and the temporal indices $\alpha$
     at various wavelengths ranging from high energy $\gamma$-rays (GeV) to radio bands,
     for the homogeneous interstellar medium case with $\delta=1$.}
    \label{fig:index1}
\end{figure}

\begin{figure*}
\centerline{\includegraphics[width=0.7\textwidth,trim= 0 0 0 0 , clip]{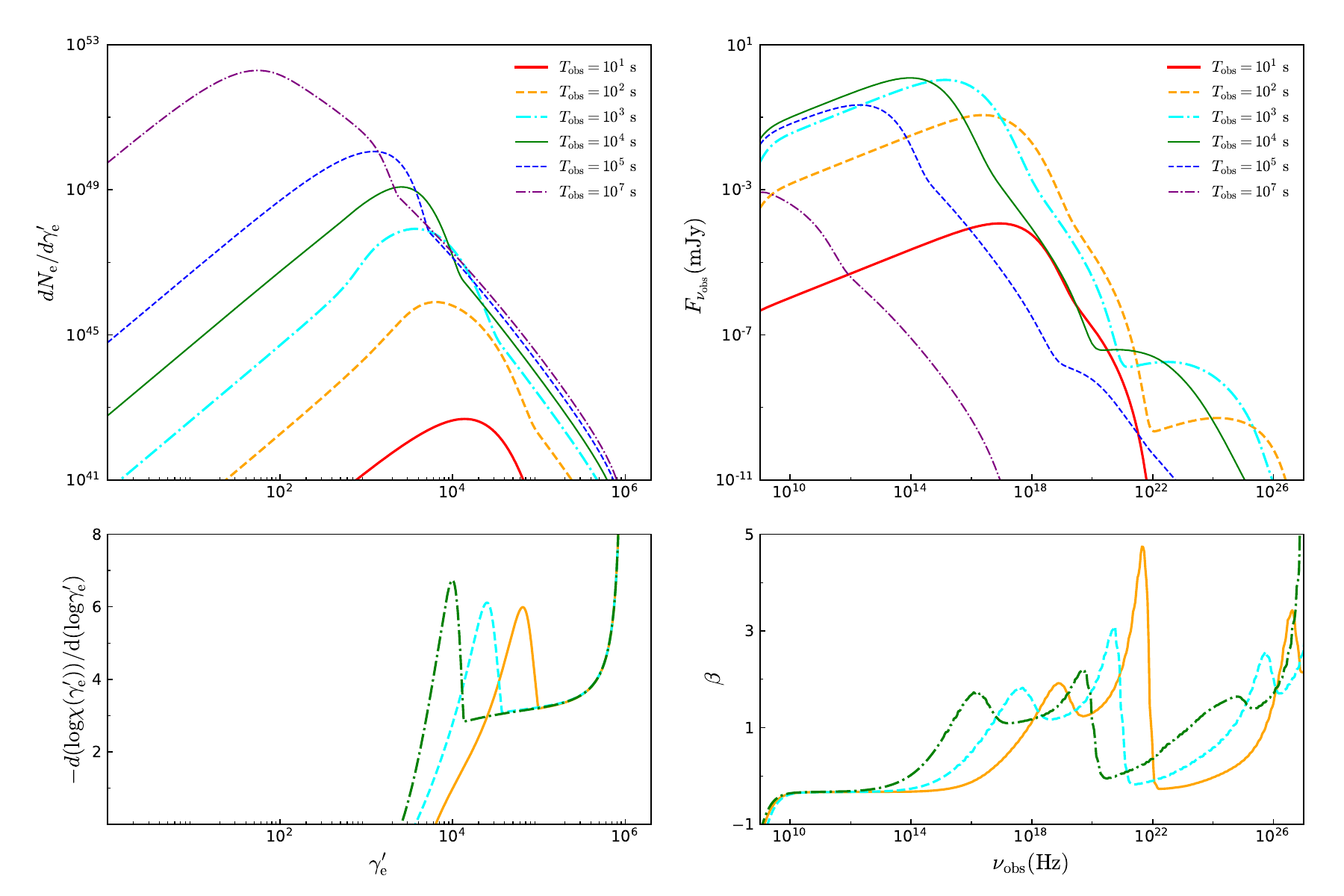}}
\caption{Same as in Figure~\ref{fig:ED1}, except that the $\delta$ parameter is taken as $\delta=0.2$.}
\label{fig:ED2}
\end{figure*}

\begin{figure}
	\centering\includegraphics[scale=.4]{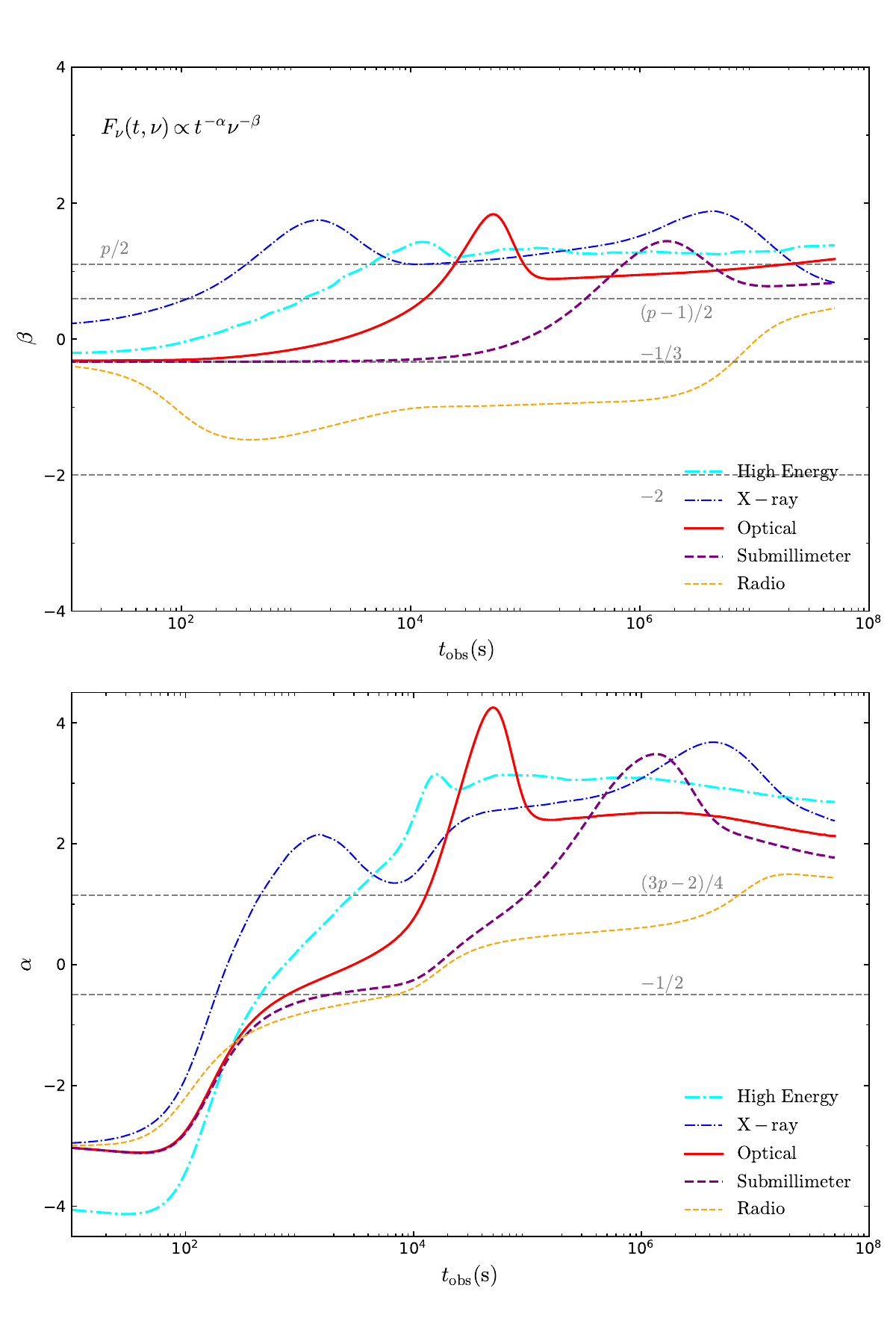}
    \caption{Same as in Figure~\ref{fig:index1}, except that the $\delta$ parameter is taken as $\delta=0.2$. }
    \label{fig:index2}
\end{figure}

\subsection{Homogeneous Interstellar Medium Case}
\label{sec:Interstellar}

In this section, we consider the case that the GRB occurs in a homogeneous interstellar
medium. The generical dynamical equations proposed by \citet{Huang00c} are adopted to
depict the deceleration of the external shock. Typical values are taken for the various
parameters involved. For example, the redshift of the GRB is taken as $z=1$; The
half-opening angle of the jet is assumed to be $\theta_{\mathrm{j}}=2^{\circ}$, with an
isotropic equivalent kinetic energy of $E=10^{53}$ erg; The initial Lorentz factor of
the jet is adopted as $\Gamma = 300$; Microphysical parameters characterizing the energy
fractions of electrons and the magnetic field are taken as $\epsilon_{\text{e}}=0.2$
and $\epsilon_{B}=0.01$, respectively; The number density of the interstellar medium
is set as $n=1.0\ \text{cm}^{-3}$. Finally, the power-law index of the electron
distribution function is $p=2.2$. Various PIC simulations indicate that the power-law
component of electrons accelerated by shocks can take a fraction of 10 -- 20 percent
of the total energy \citep{Spitkovsky08a,Spitkovsky08b,Martins09,Sironi13}.
Here, the energy fraction is defined as:
\begin{equation}
\delta = \frac{\int_{\gamma_{\mathrm{nth}}}^{\infty}
\gamma_{\mathrm{e}} Q \left(\gamma_{\mathrm{e}}, t\right)
d \gamma_{\mathrm{e}}}{\int_1^{\infty} \gamma_{\mathrm{e}}
Q\left(\gamma_{\mathrm{e}}, t\right) d \gamma_{\mathrm{e}}}.
\end{equation}
In our study, two different values are assumed for the energy fraction,
i.e. $\delta = 0.2$ and $\delta = 1$.

Figure \ref{fig:dyna1} shows the evolution of four parameters: the bulk Lorentz
factor ($\Gamma$) of the jet, the co-moving magnetic field strength ($B^{\prime}$),
the injection rate ($d N_{\mathrm{e}}/ dt_\mathrm{obs}$) of electrons, and the minimum
Lorentz factor ($\gamma_\mathrm{nth}$) of injected non-thermal electrons.
Notably, we see that $\gamma_\mathrm{nth}$ evolves significantly in the process.
At the early stage, when the observer's time ($t_{\mathrm{obs}}$) is less
than $10^{2}$ s, $\Gamma$, $B^{\prime}$, and $\gamma_\mathrm{nth}$ remain constant.
However, the injection rate ($d N_{\mathrm{e}}/ dt_\mathrm{obs}$) increases quickly by
more than two orders of magnitude during this period. After $10^{2}$ s, the jet
enters the deceleration phase so that $\Gamma$, $B^{\prime}$, and $\gamma_\mathrm{nth}$
begin to decrease continuously. The increase of $d N_{\mathrm{e}}/ dt_\mathrm{obs}$ also
becomes slower. Comparing with the case of $\delta=1$, the evolution of the bulk
Lorentz factor and the co-moving magnetic field strength is similar when $\delta=0.2$.
The presence of thermal electrons alters the distribution of accelerated electrons,
leading to a change in the injection rate and the $\gamma_{\mathrm{nth}}$ parameter.

The evolution of the electron distribution and afterglow spectrum are shown in Figure \ref{fig:ED1}.
At the early stage, the cooling of electrons is dominated by adiabatic expansion since
the jet radius is small. The energy loss rate of electrons with Lorentz factor lower than
$\gamma_{\mathrm{nth}}$ is very small, while the injection rate ($d N_{\mathrm{e}}/ dt_\mathrm{obs}$)
continues to grow. This leads to a spike in the electron distribution at $t_{\mathrm{obs}}=10$ s.
After that, with the expansion of the jet radius, synchrotron cooling becomes dominant in the
lower energy segment. Concurrently, the total number of previously injected electrons is so large
that the newly injected electrons are ineffective in reshaping the electron distribution.
As a result, low-energy electrons exhibit a broken power-law distribution at $t_{\mathrm{obs}}=10^{7}$ s.
As the electrons lose their energy through radiation, the peak frequency becomes smaller.
The GeV emission initially increases (from $t_{\mathrm{obs}}=10$ s to $t_{\mathrm{obs}}=10^{3}$ s)
due to the electron injection. However, it subsequently declines since the peak frequency of
synchrotron emission is decreasing.

The lower-left panel and lower-right panel of Figure \ref{fig:ED1} show the power-law indices of the electron distribution function and the spectral indices of the afterglow spectrum, respectively.
Note that the afterglow spectrum is comprised of two distinct components, i.e.,
the synchrotron radiation component and the inverse Compton scattering component.
Quick variation of the spectral index occurs in the high-frequency range, which
could be clearly seen in the lower-right panel as a clear bump.

The evolution of the spectral index ($\beta$) at various wavelengths (from GeV to radio bands)
is illustrated in the upper panel of Figure \ref{fig:index1}. Theoretically, the spectral index
can be derived by considering the evolution of several characteristic frequencies, which
gives \citep{Zhang06}
\begin{equation}
\beta=
\begin{cases}
-2, & \nu < \nu_\mathrm{a}, \\
-\frac{1}{3}, & \nu_\mathrm{a} < \nu < \nu_\mathrm{m}, \\
\frac{p-1}{2}, & \nu_\mathrm{m} < \nu < \nu_\mathrm{c}, \\
\frac{p}{2}, & \nu > \nu_\mathrm{c},
\end{cases}
\end{equation}
where $\nu_a$ and $\nu_c$ represent the self-absorption frequency and cooling frequency, respectively.
In the upper panel, four horizontal dashed lines corresponding to $\beta = p/2$, $\beta = (p-1)/2$,
$\beta = -1/3$, and $\beta = -2$ are also shown for a direct comparison. The peak frequency of the  synchrotron
self-Compton (SSC) emission is initially close to the GeV band, which makes the GeV spectral index be nearly zero. Later, the index gradually increases and eventually exceeds $p/2$. The spectral index in the X-ray band
evolves from $(p-1)/2$ to $p/2$ as $\nu_c$ crosses this frequency range.
Then it surpasses $p/2$ and
reaches a maximum value at $\sim 4 \times 10^6$ s, which corresponds to the bump in the spectral indices of the afterglow spectrum
as shown in Figure \ref{fig:ED1}. The evolution of the spectral indices in the optical and submillimeter
bands also has two distinct plateau phases as the characteristic frequencies traverse the corresponding
bands. However, the spectral index in the radio band exhibits a different evolution pattern due to
the  synchrotron self-absorption (SSA) effect. It first decreases, then comes to a plateau phase, and finally it rises again.
Note that since the radio frequency is slightly larger than the self-absorption frequency, the spectral
index remains above $-2$ all the time.

The evolution of the temporal indices at different frequencies is shown in the lower panel of
Figure \ref{fig:index1}. When $t_{\mathrm{obs}} \lesssim 100$ s, the number of injected electrons is
so small that the injection plays a more important role than the cooling process so that the
source term in Equation (\ref{eq:ab}) becomes dominant. In this period, the sign of $\alpha$ is
opposite to that of $\beta$. As the number of injected electrons increases,
it dominates the advection term in Equation (\ref{eq:ab}), which makes the index transit from
a negative value to a positive value.
The index in the X-ray band shows a second plateau when $\beta$ approaches $p/2$ at $t_{\mathrm{obs}} \sim 10^{3}$ s.
When $t_{\mathrm{obs}} \gtrsim 5 \times 10^{3}$ s, the peak Lorentz factor ($\gamma_{\mathrm{peak}}$,
see Figure \ref{fig:ED1}) of electrons surpasses the value of $\gamma_{\mathrm{nth}}$.
Note that the Lorentz factor of the injected electrons is typically $\gamma_{\mathrm{nth}}$
($ Q_{\mathrm{inj}} \propto \gamma_{\mathrm{nth}}^{-p}$). After that, the electron injection
declines rapidly, leading to a reduction in the contribution of the source term. As a result,
$\alpha$ enters a plateau phase. When $t_{\mathrm{obs}} \gtrsim 4 \times 10^{6}$ s, the source term
becomes negligible, leading to a bump in the evolution curve of $\alpha$, which is a counterpart
of that observed in the upper panel. The timing index of the GeV emission follows a similar evolution
to that of the X-ray band, because GeV photons are originated from X-ray photons up-scattered by
energetic electrons. In the submillimeter band, the timing index exhibits a plateau at
$\alpha = -1/2$, which corresponds to a spectral index of $\beta= -1/3$. It is followed by a
second plateau phase at $t_{\mathrm{obs}} \approx 10^{5}$ s. The timing index of radio emission also shows
a plateau at around $-1/2$. It slowly rises after $\sim2 \times 10^{4}$ s and eventually shows a second
plateau at $10^{7}$ s. Generally, the evolution of the timing index of optical afterglow is between
those of the X-ray band and the submillimeter band.

The evolution of the electron distribution and afterglow spectrum in the case
of $\delta=0.2$ is shown in Figure \ref{fig:ED2}. In the low-energy range where
the Maxwellian distribution takes on a quadratic function form, the cooling of
electrons is negligible so that they follow a power-law distribution.
The transition between the Maxwellian component and the power-law component
naturally leads to a relatively large power-law index of the electron distribution
function. Since the afterglow spectrum completely depends on the electron
distribution, we see that the spectral index near $10^{16}$ Hz increases markedly
at $t_{\mathrm{obs}} = 10^{3}$ s. In the lower-right panel of Figure \ref{fig:ED2},
there is an obvious bump-like structure at $\sim10^{19}$ Hz. It is due to the superposition
of the synchrotron radiation and SSC emission. The photons of $\sim10^{16}$ Hz can be
up-scattered by energetic electrons through the SSC process, which leads to
another bump at $\sim10^{25}$ Hz.

The evolution of the spectral and temporal indexes in different bands in the case
of $\delta=0.2$ is presented in Figure \ref{fig:index2}.
The X-ray spectral index shows a non-monotonic hard-soft-hard variation during the plateau phase observed in the case of $\delta=1$.
The second bump in the X-ray band is similar to that
in the $\delta=1$ scenario. There is also a less evident bump in the GeV range.
Both the optical and sub-millimeter spectral indices exhibit similar bumps during
the late stage. Due to the SSA effect, the evolution of the radio band spectral index
is similar to that when $\delta=1$. The lower panel of Figure \ref{fig:index2} shows
the evolution of the temporal index ($\alpha$). We see that when the spectral index
($\beta$) rises, the temporal index generally also increases. This could be an important
indication of the thermal electrons.

\begin{figure}
	\centering\includegraphics[scale=.3]{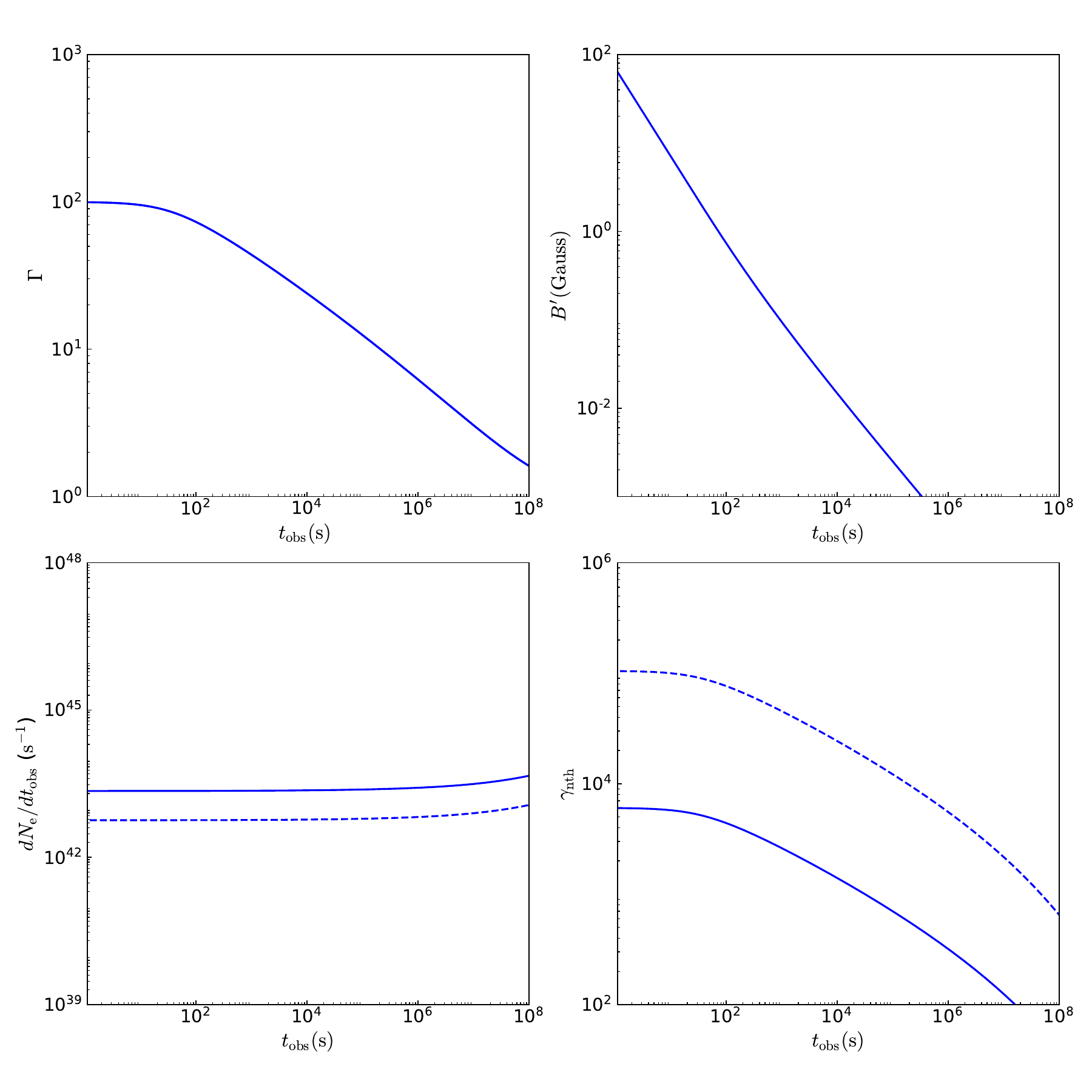}
    \caption{ Similar to Figure \ref{fig:dyna1}, but the circumburst medium is the stellar wind. }
    \label{fig:dynasw1}
\end{figure}

\begin{figure*}
\centerline{\includegraphics[width=0.7\textwidth,trim= 0 0 0 0 , clip]{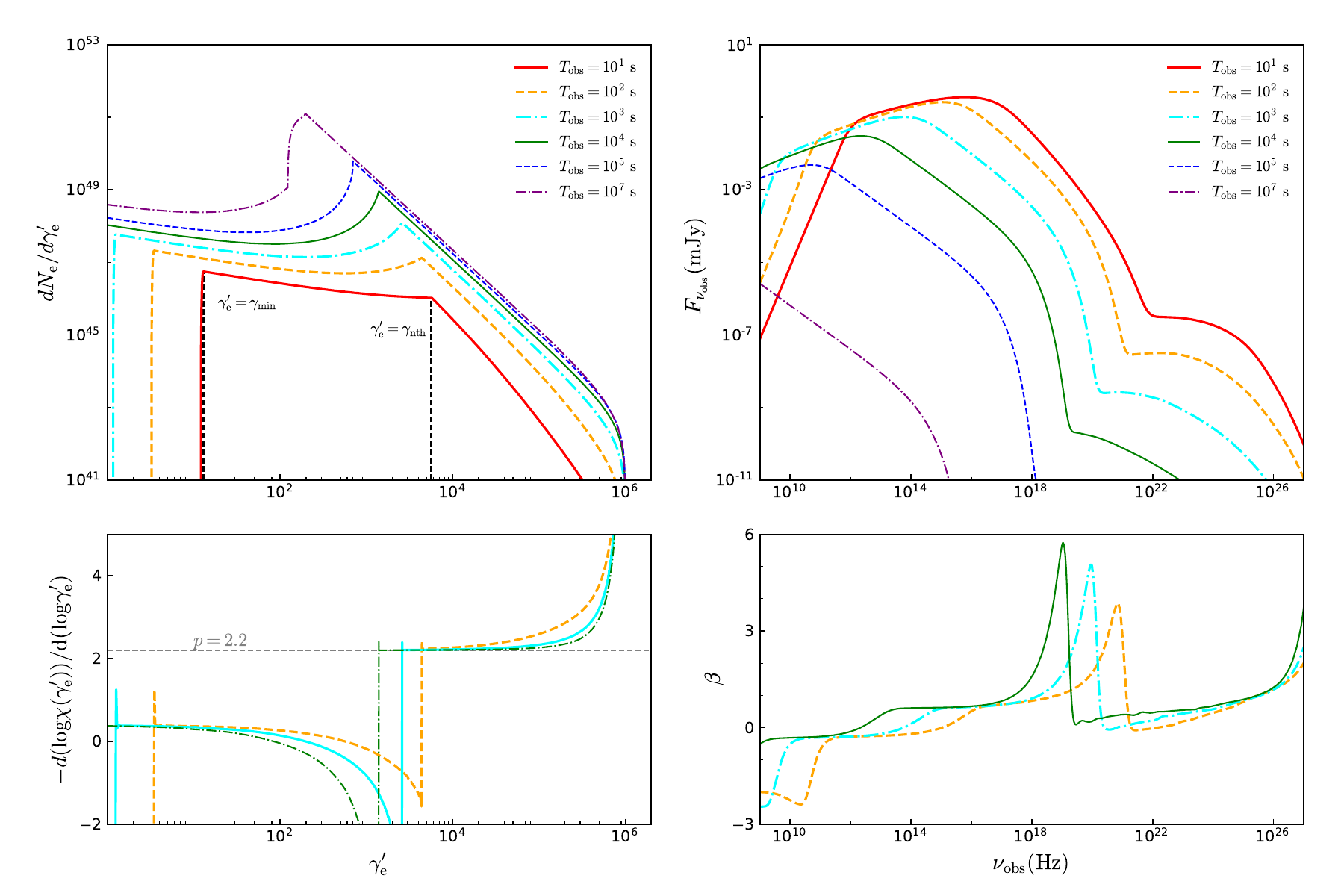}}
\caption{
   Similar to Figure \ref{fig:ED1}, but the circumburst medium is the stellar wind. The characteristic Lorentz factors of $\gamma_\mathrm{min}$ and $\gamma_\mathrm{nth}$ are marked in the upper-left panel.}
\label{fig:EDsw1}
\end{figure*}

\begin{figure}
	\centering\includegraphics[scale=.4]{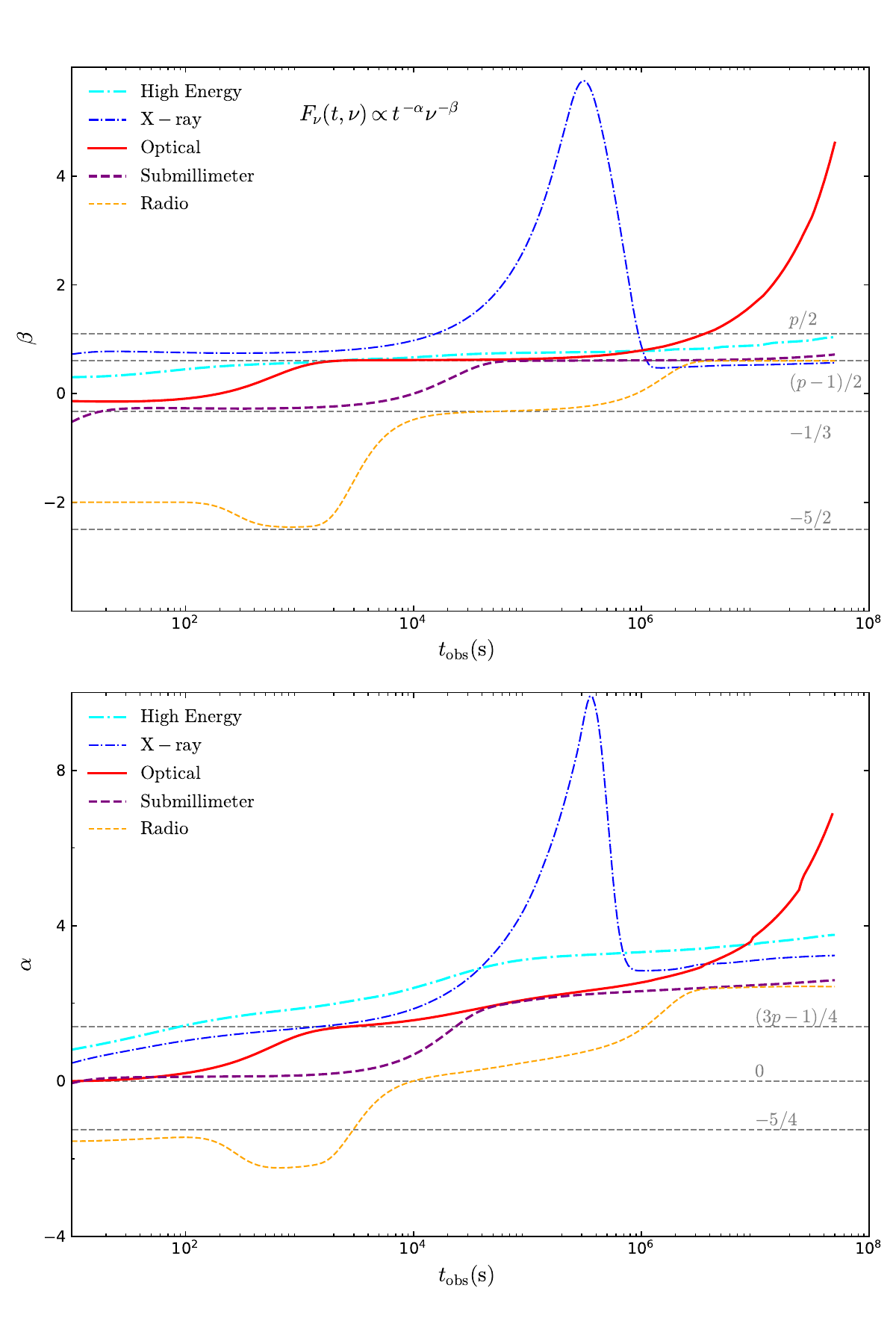}
    \caption{ Similar to Figure \ref{fig:index1}, but the circumburst medium is the stellar wind.     }
    \label{fig:indexsw1}
\end{figure}

\begin{figure*}
\centerline{\includegraphics[width=0.7\textwidth,trim= 0 0 0 0 , clip]{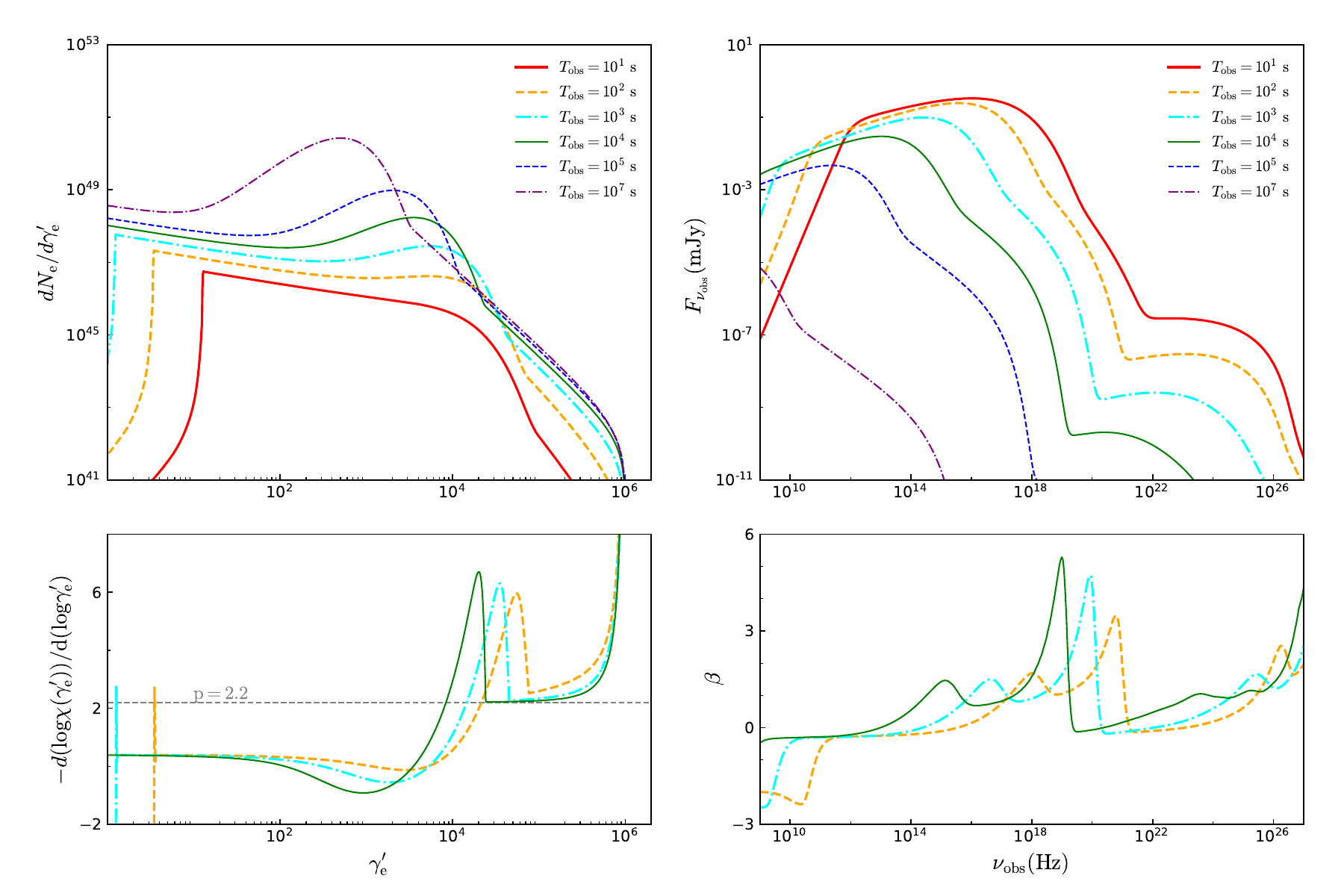}}
\caption{ Same as in Figure~\ref{fig:EDsw1}, except that the $\delta$ parameter is taken as $\delta=0.2$. }
\label{fig:EDsw2}
\end{figure*}

\begin{figure}
	\centering\includegraphics[scale=.4]{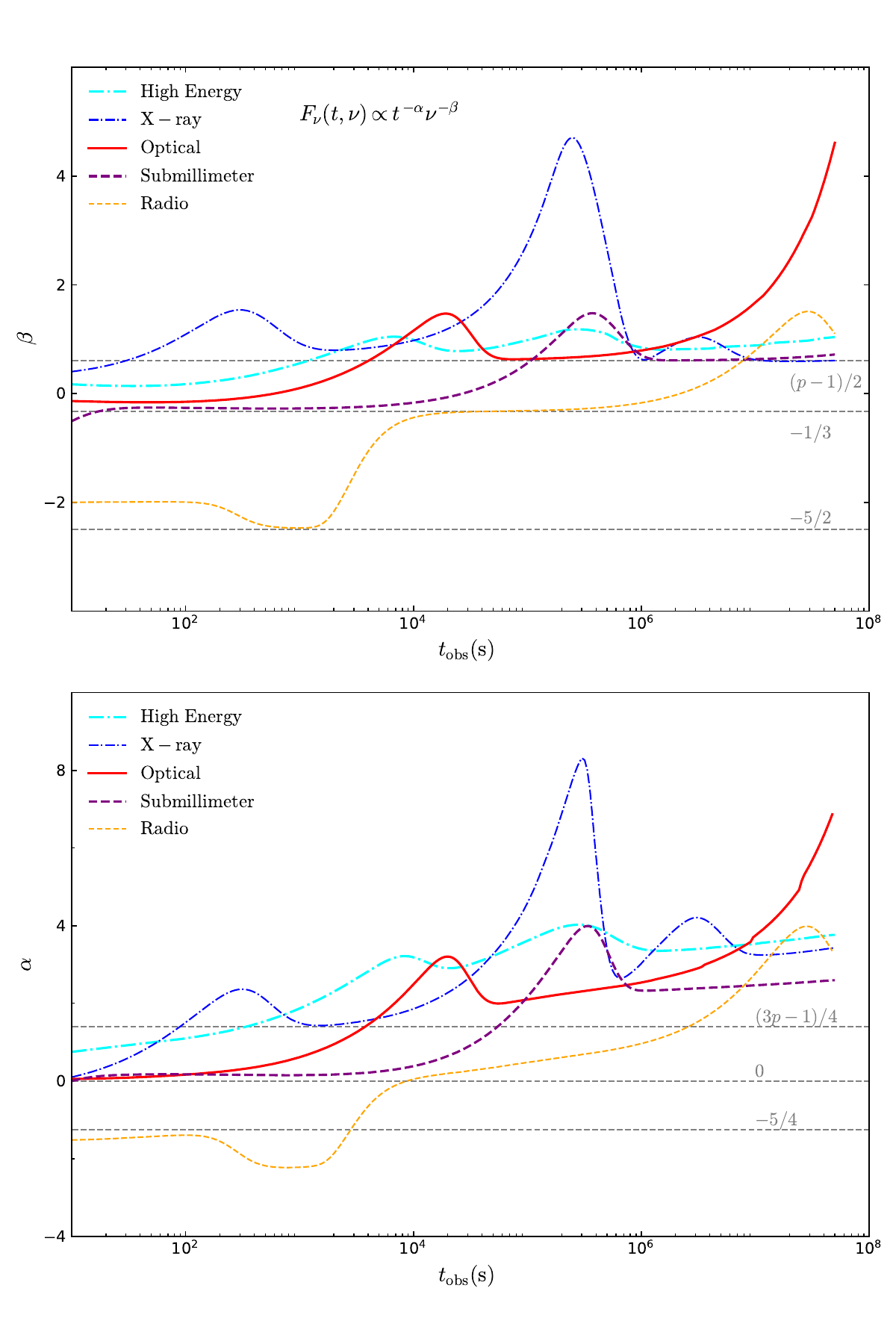}
    \caption{ Same as in Figure~\ref{fig:indexsw1}, except that the $\delta$ parameter is taken as $\delta=0.2$. }
    \label{fig:indexsw2}
\end{figure}

\subsection{Stellar Wind Medium Case}
\label{sec:StellarWind}

In this subsection, we present our numerical results on GRBs that occur in a
stellar wind environment. Again we consider two values for the $\delta$ parameter,
i.e. $\delta=0.2$ and $\delta=1$. The energy fraction of the magnetic field is
taken as $\epsilon_{B}=10^{-4}$. The density of the stellar wind is assumed to take
the form of $n(r)=3 \times 10^{35}A_{\star}r^{-2}\ \text{cm}^{-3}$, where $A_{\star}$
is a normalization constant of the wind density which is taken
as $A_{\star}=0.01$. Other parameters are the same as those taken in the
previous subsection.
The main conclusion of this subsection would not be changed for large $A_{\star}$ values.

Figure \ref{fig:dynasw1} shows the evolution of four parameters: the bulk Lorentz
factor ($\Gamma$) of the jet, the co-moving magnetic field strength ($B^{\prime}$),
the injection rate ($d N_{\mathrm{e}}/ dt_\mathrm{obs}$) of electrons, and the minimum
Lorentz factor ($\gamma_\mathrm{nth}$) of injected non-thermal electrons. Two cases
of $\delta=1$ and $\delta=0.2$ are considered in our calculations. After a
short coasting phase, the bulk Lorentz factor begins to decrease at 10 seconds.
The magnetic field decays monotonously, but its initial value is higher than that
of the homogeneous interstellar medium case. The injection rate of electrons only
varies in a narrow range. Meanwhile, the Lorentz factor of $\gamma_{\mathrm{nth}}$
also decays monotonously.

The evolution of the electron distribution function and the afterglow spectrum
is illustrated in Figure \ref{fig:EDsw1}. Since the magnetic field strength here is
larger than that of the homogeneous interstellar medium scenario, the cooling of
low-energy electrons is more pronounced than that in Figure \ref{fig:ED1},
leading to a broken power-law function for the electron distribution.
As $\gamma_{\mathrm{nth}}$ slowly decreases, the injection process begins to affect
the electron distribution and results in an upward warp, which emerges as a noticeable
bulge at around $10^7$ seconds.
Comparing with the interstellar medium case, there is a similar bump in the lower-right
panel, which shows the evolution of the spectral index of the multi-band afterglow.

The upper panel of Figure \ref{fig:indexsw1} illustrates the evolution of $\beta$.
The X-ray spectral index is initially between $p/2$ and $(p-1)/2$. It reaches a peak
value of nearly 6 at around $2 \times 10^{5}$ s, and then returns to $(p-1)/2$ later.
In the GeV band, the evolution of the spectral index is very flat. It approximately
equals to $(p-1)/2$ all the time.
The index in the optical band is negative initially. It gradually increases between
$10^2$ --- $10^3$ s and keeps to be about $(p-1)/2$ during $10^3$ --- $10^6$ s.
Then the optical index increases significantly at later stages.
The spectral index in the sub-millimeter band is initially $-1/3$. It increases
to $(p-1)/2$ at around $10^4$ s and remains constant thereafter.

Theoretically, when $\nu_c < \nu_a < \nu_m$, the spectral index should take
the values of $-2, -5/2, 1/2$, and $p/2$ successively from high frequency regime
to low frequency regime. On the other hand, when $\nu_a<\nu_c<\nu_m$, the index
should vary as $-2, -1/3, 1/2$, and $p/2$ from high to low frequency.
In the radio bands, the increasing jet radius causes a reduction in the electron
density and leads to a decrease in the SSA frequency. Furthermore, the cooling of
electrons changes their distribution. Consequently, the evolution of the radio
band spectral index becomes very complicated. In Figure \ref{fig:indexsw1}, we see
that the radio-band spectral index experiences three distinct plateau phases.
Initially, radio waves are emitted by electrons whose Lorentz factor is approximately
$\gamma_{\mathrm{min}}$ (see the upper left panel of Figure \ref{fig:EDsw1}),
leading to a spectral index of $-2$ due to the SSA process. As the magnetic field
strength declines, radio waves are emitted mainly by non-thermal electrons which
obeys a power-law distribution, but not by the monochromatic electrons
at $\gamma_{\mathrm{min}}$. The spectral index correspondingly changes to $-5/2$.
After that, radio waves will mainly come from those electrons with the Lorentz
factor concentrating at $\gamma_{\mathrm{nth}}$. By this time, the jet radius
has grown significantly, rendering the impact of SSA negligible and resulting
in a radio spectral index of $-1/3$. Finally, radio waves will come from those
electrons with the Lorentz factor larger than $\gamma_{\mathrm{nth}}$, which
essentially follow a power-law distribution
of $\frac{d N_{\mathrm{e}}}{d \gamma_{\mathrm{e}}^{\prime}}
\propto \gamma_{\mathrm{e}}'^{-p}$, then we have a radio spectral index of $(p-1)/2$.
The above evolution pattern of the radio spectral index can be clearly seen
in Figure \ref{fig:indexsw1}.

The upper panel of Figure \ref{fig:indexsw1} shows the evolution of the
temporal index $\beta$. Here, the magnetic field strength is larger than
that of the homogeneous interstellar medium scenario, and the advection
term in Equation (\ref{eq:ab}) is dominant even in the early stage.
As a result, the sign of $\beta$ is the same as that of $\alpha$.
Again, we see a similar bump in the X-ray band as that of the interstellar
medium case. In optical and radio bands, the evolution of the temporal
index shows a clear correlation with that of the spectral index when
comparing the two panels of Figure \ref{fig:indexsw1}.

The evolution of the electron distribution function and the afterglow spectrum
is illustrated in Figure \ref{fig:EDsw2} for the scenario of $\delta=0.2$.
The upper-left panel clearly shows that the electron distribution is a combination
of a power-law component and a Maxwellian component. Consequently, there is a
noticeable bulge structure in the lower-left panel which shows the electron
distribution index versus the Lorentz factor. The lower-right panel shows the
spectral index at different frequency. Again, several distinct bulges can be seen
in the plot. For example, at $t_{\mathrm{obs}} = 10^2$ s, the existence of thermal
electrons results in the first bulge at $\sim 10^{18}$ Hz. The second bulge at
$\sim 10^{21}$ Hz is caused by the overlapping of the SSC and synchrotron emissions.
Finally, since the SSC emission strongly depends on the electron distribution and
seed synchrotron photons, the first bulge (at $\sim 10^{18}$ Hz) leads to a
third bulge at $\sim 10^{26}$ Hz.

The evolution of the spectral and temporal indices is illustrated in
Figure \ref{fig:indexsw2} for the scenario of $\delta=0.2$. In the upper
panel, the bump at $t_{\mathrm{obs}}=10^2$ s in X-ray band is induced by
the bulge of the electron distribution index (see the lower-left panel of
Figure \ref{fig:EDsw2}). It could be regarded as a signature for the existence
of thermal electrons. Similarly, in optical, sub-millimeter, and radio bands,
a bump can be seen at $t_{\mathrm{obs}} \sim 2 \times 10^{4}$ s,
$4 \times 10^{5}$ s and $2\times 10^{7}$ s, respectively.
In the lower panel of Figure \ref{fig:indexsw2}, the evolution of the temporal
index shows an obvious correlation with that of the corresponding spectral index.
In each band, the temporal index exhibits similar bump-like structure as the
spectral index.

\begin{figure*}
\centerline{\includegraphics[width=0.95\textwidth,trim= 0 0 0 0 ,clip]{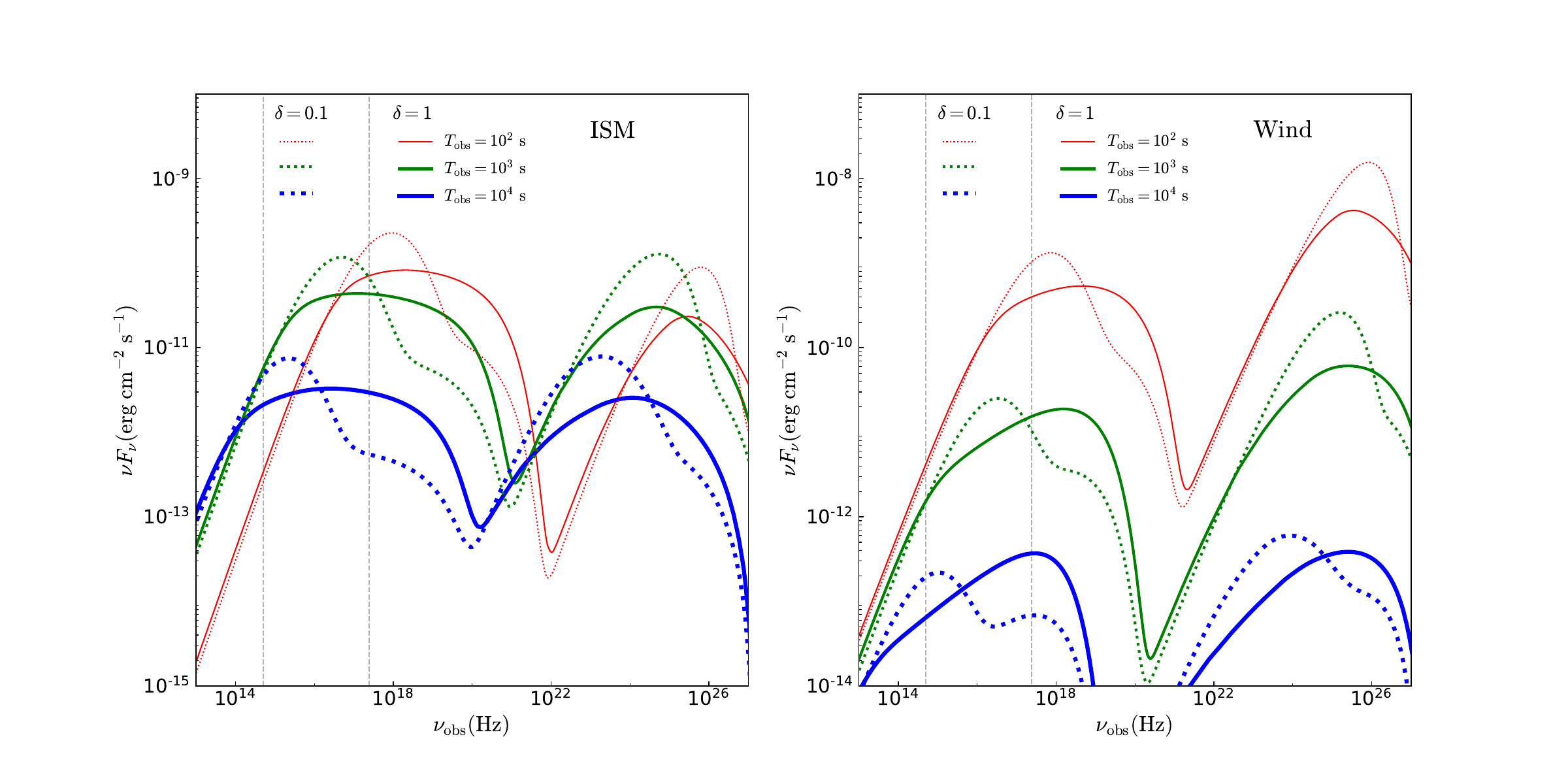}}
\caption{The afterglow spectra at three representative moments. The left panel
         shows the homogeneous interstellar medium cases and the right panel
         shows the stellar wind cases. The dotted lines correspond to $\delta=0.1$,
         and the solid lines correspond to $\delta=1$.
         Two dashed vertical lines represent the optical band and the soft X-ray band, respectively.}
\label{fig:ES}
\end{figure*}

\subsection{Afterglow Spectrum}
\label{sec:ES}

The effects of thermal electrons on the evolution of the spectral and temporal
indices have been discussed in the previous subsections. Here we further examine
the influence of thermal electrons on the afterglow spectrum. Figure \ref{fig:ES}
shows the $\nu F_{\nu}$ spectrum of the afterglow for both the homogeneous
interstellar medium scenarios and the stellar wind medium scenarios.
Two conditions of $\delta=1$ and $\delta=0.1$ are considered.
Generally, from these plots, we can see two distinct components in the spectrum, which
correspond to the synchrotron emission and the SSC emission respectively.
For the homogeneous interstellar medium case, when the effect of thermal electrons is
negligible, the spectrum of the synchrotron emission is very flat and
spans 4 -- 5 orders of magnitudes in frequency range. On the contrary, the
effect of thermal electrons mainly shows up as a peak in the low frequency region.

The spectra of the stellar wind scenarios are shown in the right panel of
Figure \ref{fig:ES}. When $\delta=1$, since the electrons follow a broken
power-law distribution, the synchrotron emission correspondingly shows a
clear break in the spectrum (also see Figure \ref{fig:EDsw2}).
However, when $\delta=0.1$, the spectrum is clearly different, which has a
bump in both the synchrotron and SSC component at $t_{\mathrm{obs}}=10^{2}$ s.
At later stages, the bump structure becomes even more prominent. It could
be regarded as a clear indication for the existence of thermal electrons.

\begin{figure*}
\centerline{\includegraphics[width=\textwidth,trim= 0 0 0 0 , clip]{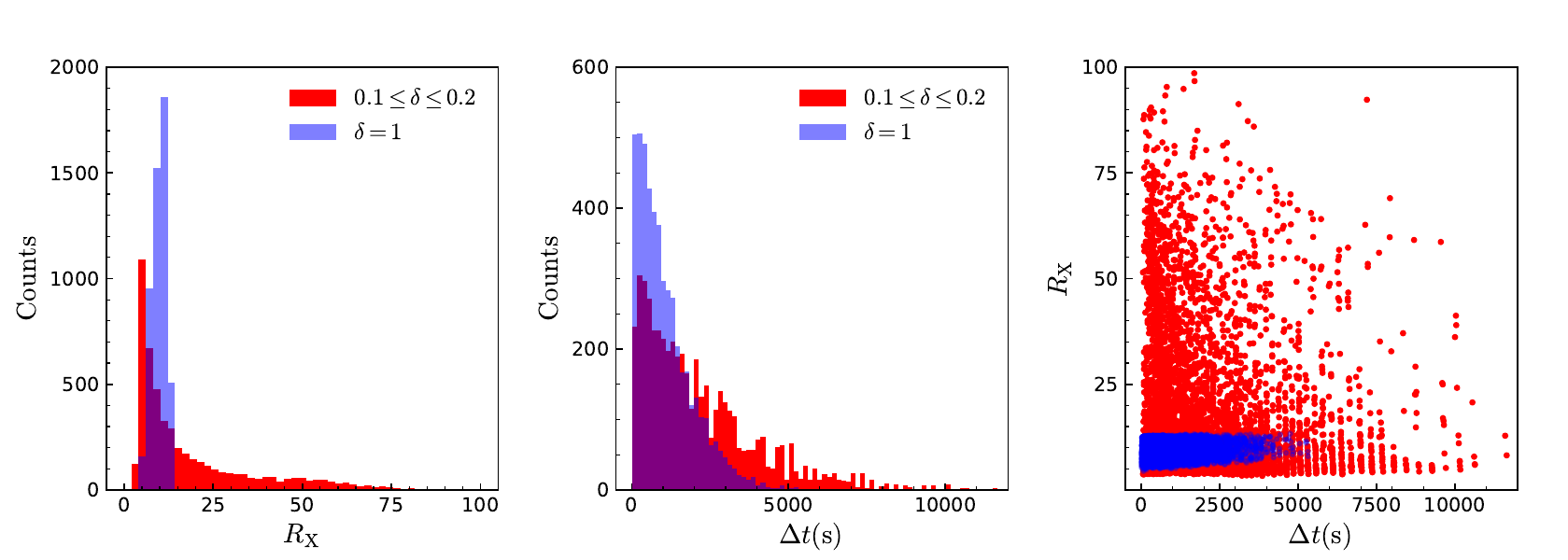}}
\caption{The distribution of simulated GRB afterglows that occur in the homogeneous interstellar medium.
The left panel shows the histograms of the peak flux ratio ($R_{\mathrm{X}}$) between soft X-ray (10 keV) and hard X-ray (100 keV) emissions for simulated GRBs.
The middle panel displays the histograms of the time delay between the peak times of soft X-ray and optical bands.
The right panel presents the distribution of the bursts on the $R_{\mathrm{X}}$-$\Delta t$ plane.
The blue dots or histograms correspond to GRBs simulated with $\delta=1$, while the red dots or histograms represent GRBs with $0.1 \leq \delta \leq 0.2$.}
\label{fig:OS}
\end{figure*}

\begin{figure*}
\centerline{\includegraphics[width=\textwidth,trim= 0 0 0 0 , clip]{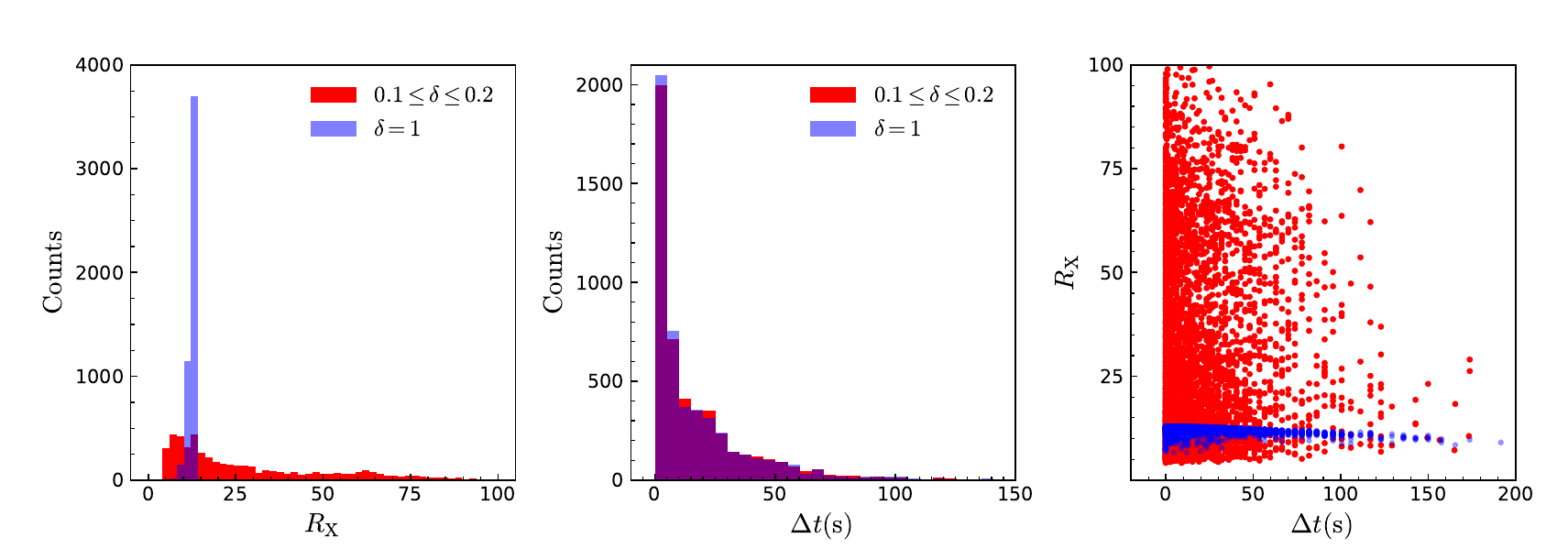}}
\caption{
Same as in Figure~\ref{fig:OS}, except that the GRBs occur in the stellar wind environment.}
\label{fig:OS2}
\end{figure*}

\section{Distribution of some observable parameters}
\label{sec:correlation}

In order to help find more credible evidence of the existence of thermal electrons through
observations, we have conducted Monte Carlo simulations to generate a large number of
GRBs. The observable parameters of their afterglows are then analyzed and their distributions
are carefully examined.
In our simulations, the redshift is assumed to be $z=1$.
Two conditions are considered for the energy fraction ($\delta$): (i) fixed as $\delta=1$; (ii) or a homogeneous distribution in the
range of $0.1 \leq \delta \leq 0.2$. The reason for the latter condition is that it is supported
by some recent PIC simulations \citep{Spitkovsky08a,Spitkovsky08b,Martins09,Sironi13}.
Ten thousand GRBs are generated under each condition for both the homogeneous interstellar medium case and the stellar wind medium case, respectively.

We mainly analyzed two parameters that can be directly measured through future
afterglow observations: the peak flux ratio ($R_{\mathrm{X}}$) between soft X-ray (10 keV) and hard
X-ray (100 keV) emissions, and the time delay ($\Delta t$) between the peak
times of soft X-ray and optical bands. These two parameters are calculated for
each simulated GRB.
Figure \ref{fig:OS} and Figure \ref{fig:OS2} show the frequency histograms for the peak flux ratio and the time delay for distinct circum-burst medium, respectively.
They also illustrate the distribution of all the mock GRBs in the $R_{\mathrm{X}}$-$\Delta t$ parameter space.

For the homogeneous interstellar medium case, parameters involved in the simulations are assumed to follow uniform distributions in the logarithmic space as: $10^{51}\ \mathrm{erg}\leq E \leq 10^{53}\ \mathrm{erg}$, $100 \leq \Gamma \leq 300$, $0.01 \leq \epsilon_{\mathrm{e}} \leq 0.5$, $10^{-3}\leq \epsilon_{B} \leq 10^{-1}$, and $0.1$ cm$^{-3} \leq n_{0} \leq 10$ cm$^{-3}$.
The corresponding simulation results are exhibited in Figure \ref{fig:OS}.
The left panel of Figure \ref{fig:OS} shows the histograms of $R_{\mathrm{X}}$ for simulated GRBs.
When there are no thermal electrons so that all the electrons follow a pure power-law distribution, a rough upper limit of $\sim 15$ can be noticed for the flux ratio.
On the contrary, when thermal electrons are present, there should be a large number of lower energy electrons.
As a result, the hard X-ray emission will be suppressed while the soft X-ray emission will be enhanced.
Consequently, $R_{\mathrm{X}}$ will be significantly increased.
In the left panel, we could see that for the pure non-thermal electron scenarios ($\delta=1$, blue histograms), all GRBs have a flux ratio roughly less than $\sim 15$.
On the other hand, for the hybrid electron scenarios with both thermal and non-thermal electrons ($0.1 \leq \delta \leq 0.2$), a significant number of GRBs (about $\sim 40\%$) have a flux ratio approximately larger than $\sim 15$.

The middle panel shows the histograms of the time delay for simulated GRBs.
For the pure non-thermal electron scenarios ($\delta=1$, blue histograms), the upper limit for $\Delta t$ is approximately $\sim 5000$ s.
As the peak frequency of the afterglow spectrum crosses the frequency of optical emission, the optical emission will reach its peak flux.
However, when other model parameters are consistent, the initial peak Lorentz factor ($\gamma_{\mathrm{peak}}$) of the electron distribution for the hybrid electron scenario is always larger than that for the pure non-thermal electron scenario.
Then the optical emission for the hybrid electron scenario will reach its peak flux at a later time.
Consequently, for the hybrid electron scenarios, the time delay can exceed $\sim 5000$ s (with a percentage of $\sim 10\%$).

The right panel shows the distribution of the bursts on the $R_{\mathrm{X}}$-$\Delta t$ plane.
The presence of thermal electrons significantly raise the upper limits of both $R_{\mathrm{X}}$ and $\Delta t$.
Then the distribution of the GRB afterglows with thermal electrons is more scattered in the $R_{\mathrm{X}}$-$\Delta t$ plane.

For the stellar wind medium case, parameters involved in the simulations are assumed to follow uniform distributions as: $10^{51}\ \mathrm{erg}\leq E \leq 10^{53}\ \mathrm{erg}$, $100 \leq \Gamma \leq 300$, $0.01 \leq \epsilon_{\mathrm{e}} \leq 0.5$, $10^{-4}\leq \epsilon_{B} \leq 10^{-2}$, and $10^{-2} \leq A_{\star} \leq 10^0$.
The corresponding simulation results are exhibited in Figure \ref{fig:OS2}.
When all the electrons follow a pure power-law distribution, a rough upper limit of $\sim 15$ can also be found for the flux ratio in the left panel of Figure \ref{fig:OS2}.
However, the X-ray peak flux of GRBs that occur in the stellar wind environment appears in the very early phase due to the density of the stellar wind following a form of $n(r) \propto r^{-2}$.
As shown in the middle panel of Figure \ref{fig:OS2}, the hybrid electron scenario shows a similar distribution of $\Delta t$ with that of the pure non-thermal electron scenario, and $\Delta t$ of most mock GRBs is less than 10 s.
The presence of thermal electrons significantly raises the upper limits of $R_{\mathrm{X}}$, therefore the distribution of the GRB afterglows with thermal electrons is still more scattered in the $R_{\mathrm{X}}-\Delta_t$ plane of Figure \ref{fig:OS2}.

\section{Observability}
\label{sec:observation}

\begin{figure}
	\centering\includegraphics[scale=.4]{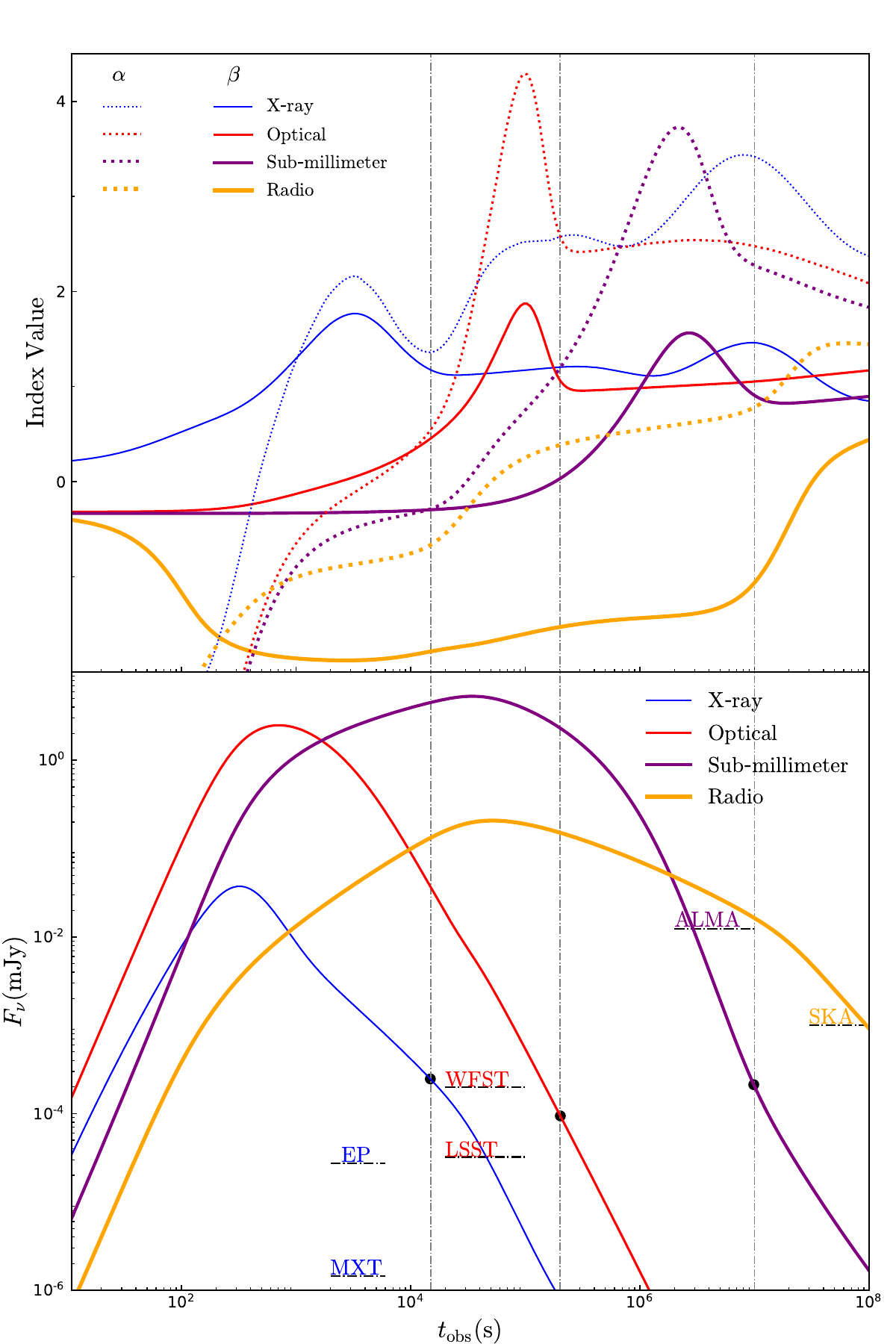}
    \caption{
    The evolution of the spectral and temporal indices at various wavelengths ranging from X-ray to radio bands, as well as the corresponding light curves,
    for the homogeneous interstellar medium case with $\delta=0.2$.
    The end times of the spectral indices decline for three bands are shown by dashed vertical lines, and the corresponding flux of these bands is marked by black dots.
    The sensitivities of several telescopes/detectors are also marked as a direct comparison, including the MXT, EP, LSST, WFST, ALMA, and SKA.
    }
    \label{fig:schematic}
\end{figure}

\begin{figure*}
\centering
    \subfloat{\includegraphics[width=9cm,height=9cm]{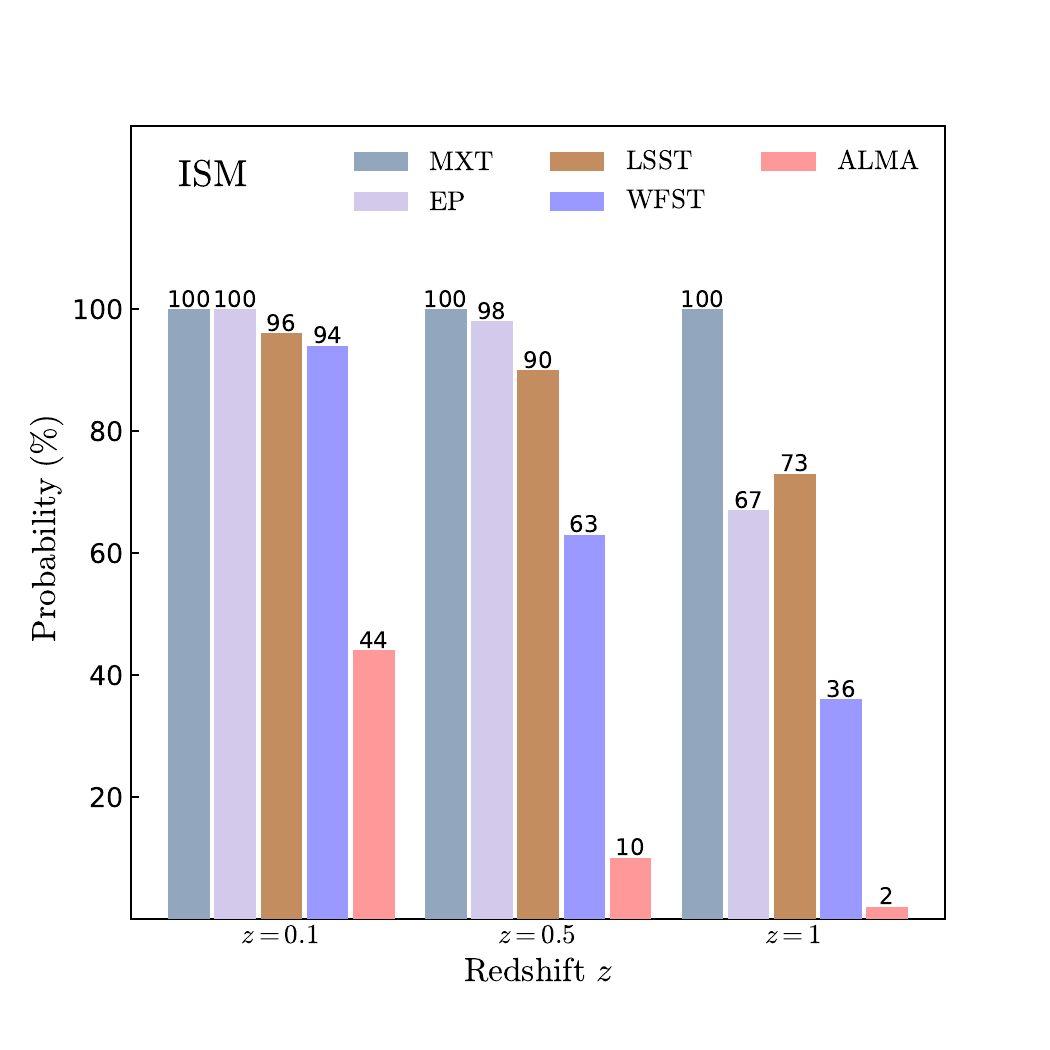}}
    \subfloat{\includegraphics[width=9cm,height=9cm]{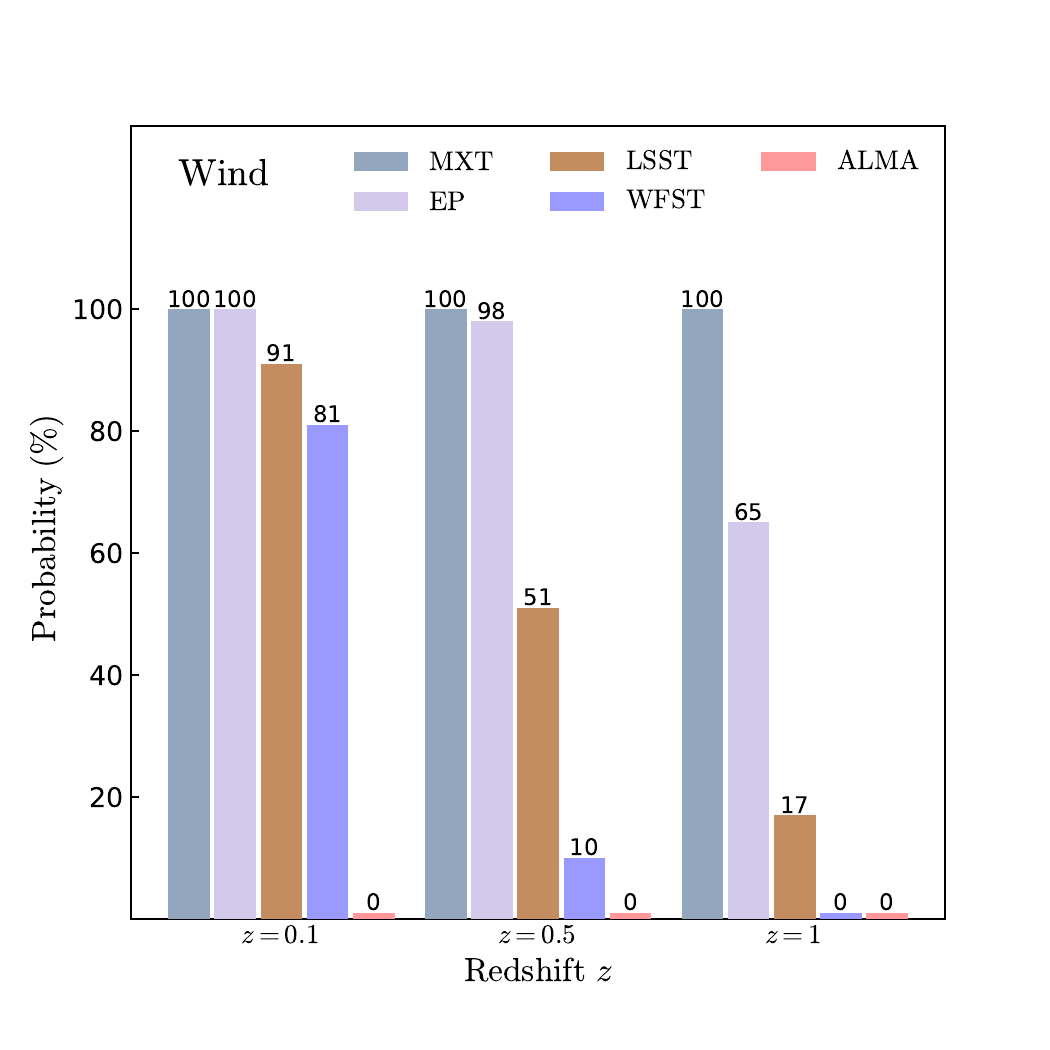}}
    \caption{The optimistic detection probability of several telescopes/detectors, including the MXT, EP, LSST, WFST, and ALMA, for detecting the thermal electron signals with spectral and timing indices in the ideal observation situation.
    The left panel shows the homogeneous interstellar medium cases and the right panel shows the stellar wind cases. }
    \label{fig:singleband}
\label{groupa}
\end{figure*}
\subsection{Single-band observability}
\label{sec:so}

As shown in Section \ref{sec:calculations}, the simultaneous occurrence
of bumps in both the spectral and temporal indices can serve as an indication
of the presence of thermal electrons. However, measuring these indices requires
that the radiation flux should be higher than the sensitivity of the detectors.
In this subsection, we discuss the single-band observability of GRB afterglows by various telescopes/detectors in the ideal observation situation.

Figure \ref{fig:schematic} shows the evolution of the spectral and temporal indices as well as the corresponding muti-band light curves.
The sensitivities of several telescopes/detectors are also marked for a direct comparison, including the Microchannel X-ray Telescope (MXT) on board the Space Variable Objects Monitor (SVOM), Einstein Probe (EP), Wide Field Survey Telescope (WFST), Vera Rubin Observatory Legacy Survey of Space and Time (LSST), Atacama Large Millimeter/submillimeter Array (ALMA), and Square Kilometre Array (SKA).
Note that the exposure time is taken as 100 s for MXT, $10^{3}$ s for EP, 3 hr for LSST and WFST, 4 hr for ALMA, and 1 hr for SKA.

As is shown in Figure \ref{fig:schematic}, in order to observe the whole thermal-electron signal with the spectral and timing indices, the detection threshold of the detector must be lower than the radiation flux before the spectral indice decline ends.
However, due to the cooling of electrons, the appearance time of the thermal-electron signals in the lower energy bands is late.
The thermal-electron signals did not appear in the spectral/timing indices of the radio band when $t_{\mathrm{obs}} \sim 10^{8}$ s, although the radio flux has been below the detection threshold of the SKA.
It is hard to identify the signal of thermal electrons from the radio band.

To obtain the detection probability for identifying thermal electrons by observing non-monotonic changes in the spectral and temporal indices in a single band,
we ran 10000 Monte Carlo simulations for various redshifts (including $z = 0.1$, $z = 0.5$, and $z = 1$) and different circumburst medium (the homogeneous interstellar and stellar wind media).
The model parameters used are identical to those in Section \ref{sec:correlation}.

The probability of detecting thermal electron signals by different detectors is displayed in the Figure \ref{fig:singleband}.
For the homogeneous interstellar medium case (see the left panel of Figure \ref{fig:singleband}),
the optimistic probabilities of the MXT, EP, LSST, WFST and ALMA identifying the thermal electron signal from a burst with a redshift of $z\sim1$ are 100\%, 67\%, 73\%, 36\% and 2\%.

Due to the rapid dissipation of the external shock in the stellar wind medium, the radiation flux is often below the detector's detection threshold before the thermal electron signal appears.
As is shown in the right panel of Figure \ref{fig:singleband},
the optimistic probabilities of the MXT, EP and LSST identifying the thermal electron signal from a burst with a redshift of $z\sim1$ are 100\%, 65\% and 17\%, respectively, while the detection probabilities of the WFST and ALMA for the same GRB population are zero.
The WFST has an optimistic probability of 81\% for identifying the thermal electrons from the bursts with a redshift of $z\sim0.1$.
\subsection{Muti-band observability}
\label{sec:mo}

\begin{figure*}
\centerline{\includegraphics[width=\textwidth,trim= 0 0 0 0 , clip]{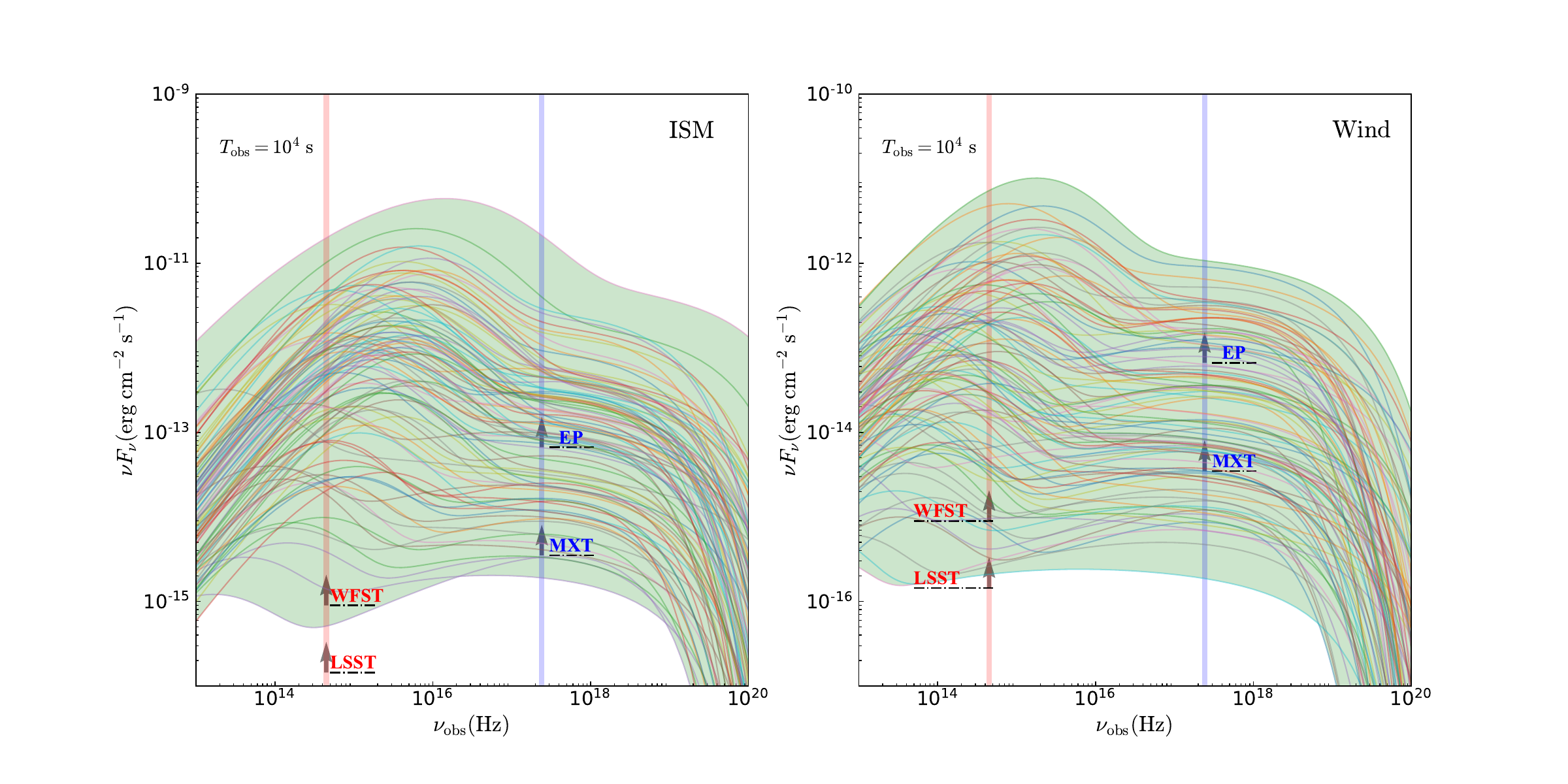}}
\caption{The Monte Carlo simulation results of the afterglow spectrum.
         The redshift is set as $z=1$, and the $\delta$ parameter is taken as $0.1\leq\delta\leq0.2$.
         The left panel shows the homogeneous interstellar medium cases and the right panel shows the stellar wind cases.
         The sensitivities of several telescopes/detectors are also marked as a direct comparison, including the MXT, EP, LSST, and WFST.
         The green shading on the graph indicates the distribution area of the afterglow spectrum.
         Two transparent solid lines represent the optical band and the soft X-ray band, respectively.
         }
\label{fig:MO}
\end{figure*}

\begin{figure}
	\centering\includegraphics[scale=.5]{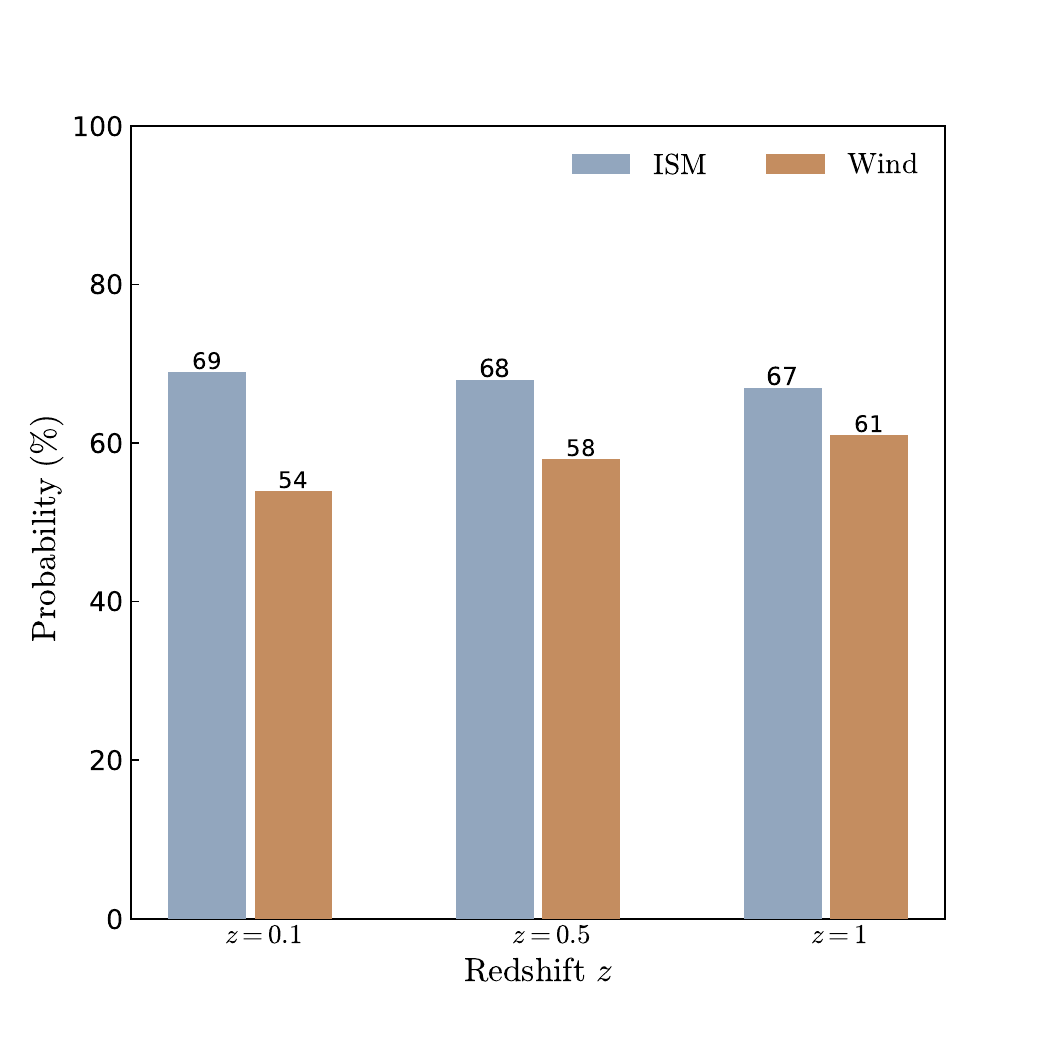}
    \caption{
    The optimistic detection probability of synergy observation in the optical (with LSST) and X-ray bands (with MXT).
    }
    \label{fig:MOp}
\end{figure}

There may be some challenges in identifying thermal electrons by observing non-monotonic changes in the spectral and temporal indices in a single band.
For example, consecutive high-cadence and high-accuracy photometry are necessary to capture the non-monotonic variation of the temporal indices.
However, there is another plausible method that identifies thermal electrons from variations of the afterglow spectrum's structure.
As is shown in Figure \ref{fig:ES}, when $t_\mathrm{obs}=10^{4}$ s, for the homogeneous interstellar medium case, the presence of thermal electrons causes the optical flux to exceed that of the soft X-ray, even though the optical flux is significantly lower in their absence; for the stellar wind case, the effect of thermal electrons makes the optical flux to be much higher than that of the soft X-ray, while it is significantly lower in the absence of thermal electrons.
Monte Carlo simulations reveal that, at the observer time of $\sim 10^{4}$ s, the probability that the presence of thermal electrons causes the optical flux to exceed that of the soft X-ray is greater than 60\%, for both the homogeneous interstellar medium and the stellar wind medium cases.
The optical flux excess phenomenon could never appear in the absence of thermal electrons regardless of the circum-burst environment of either the homogeneous interstellar medium or the stellar wind medium.
Thus, the ratio of the optical flux to that of the soft X-ray can be a indicator of the thermal-electron presence, and it could be derived by synergy observation in the optical (with LSST) and X-ray bands (with MXT).

To identify thermal electrons by synergy observation in the optical and X-ray bands, the optical and soft X-ray fluxes in the afterglow spectrum are required to exceed the detection threshold of LSST and MXT, respectively.
Moreover, if the optical flux surpasses the flux of soft X-ray, the presence of thermal electrons can be confirmed.
However, the flux ratio of between optical and soft X-ray bands in the afterglow spectrum may still be lower than 1 even if thermal electrons are present.
In this scenario, the optical flux is much smaller than that of the soft X-ray when thermal electrons are absent, thus the optical flux enhanced by the presence of thermal electrons remains lower than that of the soft X-ray.
The judgment for the presence of thermal electrons is set as the flux ratio is greater than 2 for good operability, although the ratio exceeding 1 is enough.

To obtain the detection probability of identifying thermal electrons by synergy observation in these two bands,
10000 Monte Carlo simulations are conducted for various redshifts (including $z = 0.1$, $z = 0.5$, and $z = 1$) and different circumburst mediums (the homogeneous interstellar and stellar wind media).
The model parameters used are identical to those in Section \ref{sec:correlation}.

The afterglow energy spectra of mock GRBs with a redshift of $z=1$ are depicted in Fig \ref{fig:MO}.
For most bursts, their optical and X-ray fluxes simultaneously exceed the detection thresholds of the LSST and the MXT, with the observation time set to $\sim 10^{4}$ s.
Consequently, the optimistic probabilities of detecting the thermal electron signal from a burst through synergy observation are primarily influenced by the likelihood that the optical flux surpasses the X-ray flux in $\nu F_\nu$ spectrum.
The optimistic detection probability of synergy observation in the optical (with LSST) and X-ray bands (with MXT) for detecting the thermal electron signals are shown in Figure \ref{fig:MOp}.
The optimistic probability of identifying thermal electrons from a burst with a redshift of $z\sim1$, occurring in either a homogeneous interstellar medium or a stellar wind medium, is more than 60\%.
\section{Conclusions and Discussion}
\label{sec:conclusions}

In this study, we investigate the effects of thermal electrons on GRB afterglows.
Two kinds of circum-burst medium are considered, the homogeneous interstellar medium
and the stellar wind medium. An analytical expression that connects the spectral index
and the temporal index is derived based on a simplified assumption that each electron emits
photons mainly at a particular frequency through synchrotron radiation. It is found that
a positive correlation exists between the two indices due to the cooling of electrons,
which is independent on the detailed radiation mechanism. Multi-band afterglows (from
X-ray to radio waves) are calculated numerically, which reveal that there exists a
simultaneous bump in the evolution of both the temporal and spectral index when thermal electrons
are present. The bump can serve as a crucial indicator of the presence of thermal
electrons. The presence of thermal electrons also alter the afterglow spectrum.
In the homogeneous interstellar medium scenarios, the synchrotron spectrum
exhibits a prominent bump when thermal electrons are present.
In the stellar wind medium scenarios, the synchrotron spectrum exhibits a distinctive
hump-like profile due to the thermal electrons.

Monte Carlo simulations are conducted to reveal the characteristics of GRBs in some
interesting observable parameters, i.e., the peak flux ratio between soft X-rays and
hard X-rays, and the time delay between the peak times of soft X-rays and optical
photons.

For bursts that occur in the homogeneous interstellar medium, when there are only non-thermal electrons, the peak flux ratio is generally less than $\sim 15$ and the time delay is less than $\sim 5000$ s.
However, when thermal electrons are present, the peak flux ratio can exceed $\sim 15$, and the time delay can be larger than $\sim 5000$ s.
A flux ratio exceeding $\sim 15$ or a time delay exceeding $\sim 5000$ s could be regarded as firm evidence indicating that thermal electrons are present in the shock.
For bursts that appear in the stellar wind environment,  when all the electrons follow a pure power-law distribution, a rough upper limit of $\sim$15 can also be found for the flux ratio.
When thermal electrons are present, the peak flux ratio will surpass $\sim 15$.
However, no evident difference exhibited in the distribution of $R_{\mathrm{X}}$ between the hybrid electron scenario and pure non-thermal electron scenario could be found.

For the ideal observation situation, the observability of the thermal electron signatures is discussed in context of typical telescopes/detectors.
In the homogeneous interstellar medium cases, the MXT, EP, LSST, and WFST will be able to record the bump-like signatures of the spectral and temporal indices from bursts with a redshift of $z \sim 1$.
The ALMA can identify the thermal electrons from the bursts with a redshift of $z \sim 0.5$.
In the stellar wind medium cases, the rapid decay of the afterglow generally makes it difficult to record the bump structure of the two indices.
Owing to its excellent detection performance, the MXT is able to identify thermal electrons from bursts with a redshift of $z \sim 1$, while the EP and LSST is capable of identification of thermal electrons from the same GRB population.
The WFST can identify the thermal electrons from the bursts with a redshift of $z \sim 0.1$.
The presence of thermal electrons can also be identified from the afterglow spectrum.
For both the homogeneous interstellar medium and stellar wind medium cases, the thermal electron signatures can be diagnosed by synergy observation in the optical (with the LSST) and X-ray bands (with the MXT) from burst with a redshift of $z \sim 1$.

Identifying thermal electrons through observations still poses
challenges, as it requires synchronous acquisition of multi-band data.
In fact, different instruments/detectors usually perform observations
at different times after the burst and at different cadences.
Then obtained data may even be inconsistent with each other due to
systematic difference \citep{Gobat23}.
So far, most wide-field telescopes, including WFST and LSST, can
only image the observed sky area in very limited bands during a single
exposure. It is not easy to get the temporal and spectral indices even
in optical bands. In this aspect, the Multi-channel Photometric Survey
Telescope (Mephisto) may be helpful. It has a typical $5\sigma$ r-band
depth of 22.37 mag for a 40 s exposure, and can simultaneously measure
the flux in three bands ($u, g, i$ or $v, r, z$). Both the temporal and
spectral indices could be derived from these observations.

In Section \ref{sec:correlation}, the observable characteristic of thermal electrons are shown in the peak flux ratio between soft and hard X-rays.
However, the observable parameters are hard to obtain in the real GRBs.
GRBs' prompt emission tail, which results from the curvature effect, may join the onset of the hard X-ray in the rising afterglow emission \citep{Barthelmy05b}.
Its temporal decay slope is typically in the range of $\sim$-3 to $\sim$-10.
Thus, there may be two distinct components in the 100 keV emission of early X-ray afterglows.
When the component from the external shock is not dominant, the thermal-electron signal exhibiting in the flux ratio of $R_X$ can not be distinguished.
However, if the temporal decay slope is very steep, we will have a chance to diagnose thermal electrons from the flux ratio of $R_X$.

Note that our simulations are focused on the emission from external shocks and do not include the additional energy injection from the long-lasting central engine.
The conclusion about the time delay between the peak times of soft X-ray and optical bands, can not apply to the burst with a plateau, which is beyond the scope of the current work.
In future work, we will investigate the identification of thermal electrons in the afterglows accompanied by additional energy injection.

In this work, we assume a continuous distribution of thermal and non-thermal populations in Equation (\ref{eq:distribution}).
This assumption is supported by recent PIC simulations \citep{Spitkovsky08a, Spitkovsky08b, Martins09, Sironi13, Park15, Crumley19}.
However, as discussed by \citet{Ressler17}, inefficient shock heating of thermal electrons can lead to a distinct separation between the thermal and nonthermal components.
The synchrotron emission spectrum of an electron group is broad so that the total spectrum of these two populations is still continuous if their energy separation is not too large.
In this scenario, the main conclusion of our article will not change too much.

In reality, a number of GRBs may be missed by X-ray/gamma-ray detectors on board the satellites.
Even if a GRB is triggered by these detectors, the wide-field telescope may not be able to perform timely and effective follow-up observations, due to its observatory location, weather conditions, and its own sky survey projects.
Thus, the real detection probability for detectors/telescopes of identifying thermal electrons from afterglows will be less than the simulation results.

In our Monte Carlo simulations, we assume that all the dynamic parameters
follow uniform distributions in the logarithmic space within particular ranges. It may differ from
realistic cases. The distribution of measurable afterglow parameters
could be influenced by the distribution of dynamic parameters.
Further studies on these effects need to be conducted in the future.

The acceleration mechanism of particles in relativistic shocks is still
under debate. Probing thermal electrons in GRB afterglows may provide
vital clues for understanding the particle acceleration mechanism.
Actually, there are some strange deviations from the analytical results
of spectral indices in some GRB afterglows. The contribution of thermal
electrons may account for these deviations \citep{Zhang07, Giannios09, Wang15b}.
On the other hand, the relation between the spectral and temporal indices
is related to the cooling of electrons. The observation and investigation
of both the spectral and temporal indices may help reveal the detailed
cooling process of electrons.

\begin{acknowledgments}

This study is partially supported by the National Natural Science Foundation of China (grant Nos. 12273113, 12321003, 12233002, 12393812, 12393813),
the National SKA Program of China (grant Nos. 2022SKA0130100, 2020SKA0120300), the International Partnership Program of Chinese Academy of Sciences for Grand Challenges (114332KYSB20210018), the National Key R\&D Program of China (2021YFA0718500).
Jin-Jun Geng acknowledges support from the Youth Innovation Promotion Association (2023331).
Hao-Xuan Gao acknowledges support from Jiangsu Funding Program for Excellent Postdoctoral Talent.
Yong-Feng Huang also acknowledges the support from the Xinjiang Tianchi Program.
\end{acknowledgments}

\appendix
\restartappendixnumbering

\section{The equation for spectral and temporal indices}
\label{apdx:Eqder}

The number of electrons is conserved in the cooling process, thus electrons in the interval of $[\gamma_{\mathrm{e}}^{\prime},
\gamma_{\mathrm{e}}^{\prime} + d \gamma_{\mathrm{e}}^{\prime}]$ could come from the source
injection of $ Q\left(\gamma_{\mathrm{e}}^{\prime}, t^{\prime}\right) dt^{\prime}
d \gamma_{\mathrm{e}}^{\prime}$, or from the particle flow of electrons in the interval
of $[\gamma_{\mathrm{e}}^{\prime}+\dot{\gamma_{\mathrm{e}}^{\prime}}dt^{\prime}, \gamma_{\mathrm{e}}^{\prime} +
\dot{\gamma_{\mathrm{e}}^{\prime}}dt^{\prime}+d \gamma_{\mathrm{e}}^{\prime}]$ after a short time
of $dt^{\prime}$. As a result, the numbers of electrons in the interval of $[\gamma_{\mathrm{e}}^{\prime},\gamma_{\mathrm{e}}^{\prime} + d \gamma_{\mathrm{e}}^{\prime}]$ can be expressed as
\begin{equation}
f(\gamma_{\mathrm{e}}^{\prime},t^{\prime}) d \gamma_{\mathrm{e}}^{\prime}
= f(\gamma_{\mathrm{e}}^{\prime} + \dot{\gamma_{\mathrm{e}}^{\prime}} dt^{\prime}, t^{\prime}-dt^{\prime})
[\frac{\partial \dot{\gamma_{\mathrm{e}}^{\prime}} }{\partial \gamma_{\mathrm{e}}^{\prime}}
d\gamma_{\mathrm{e}}^{\prime} dt^{\prime}+d \gamma_{\mathrm{e}}^{\prime}]
+ Q d \gamma_{\mathrm{e}}^{\prime}dt^{\prime},
\label{eq:fe}
\end{equation}
where $f(\gamma_{\mathrm{e}}^{\prime},t^{\prime})$ is the distribution of electrons and $
\dot{\gamma_{\mathrm{e}}^{\prime}}$ is the total cooling rate of electrons.
The synchrotron radiation power of electron population in the interval of $[\gamma_{\mathrm{e}}^{\prime},\gamma_{\mathrm{e}}^{\prime} + d \gamma_{\mathrm{e}}^{\prime}]$ at frequency $\nu^{\prime}$ has a similar expression of 
\begin{equation}
P^{\prime}(\nu^{\prime},t^{\prime})=P^{\prime}(\nu^{\prime} +
d\nu^{\prime},t^{\prime}-dt^{\prime})(\frac{\partial \dot{\gamma_{\mathrm{e}}^{\prime}}}
{\partial \gamma_{\mathrm{e}}^{\prime}} dt^{\prime}+1)+\hat{Q}dt^{\prime},
\label{eq:Pv}
\end{equation}
where $P^{\prime}(\nu^{\prime},t^{\prime})d\nu^{\prime}=\frac{{B^{\prime}}^{2}}{6 \pi}{\sigma}_{T}c {\gamma_{\mathrm{e}}^{\prime}}^{2}f(\gamma_{\mathrm{e}}^{\prime},t^{\prime})d \gamma_{\mathrm{e}}^{\prime}$, $\hat{Q}=\frac{{B^{\prime}}^{2}}{6 \pi}{\sigma}_{T}c {\gamma_{\mathrm{e}}^{\prime}}^{2} \frac{d\gamma_{\mathrm{e}}^{\prime}}{d\nu^{\prime}}Q, $ and $\nu^{\prime}=\frac{3 q_{\mathrm{e}} B^{\prime}{\gamma_{\mathrm{e}}^{\prime}}^{2}}{4 \pi m_{\mathrm{e}}c}$.

Note that the observed flux is determined by the synchrotron radiation power, which has a form of
\begin{equation}
F_{\nu_{\mathrm{obs}}}=\frac{(1+z) \Gamma P^{\prime}\left(\nu^{\prime}\left(\nu_{\mathrm{obs}}\right)\right)}{4 \pi D_L^2},
\label{eq:Fv}
\end{equation}
where $\nu^{\prime}=(1+z)\nu_{\mathrm{obs}}/\mathcal{D}$. Then, we can get
\begin{equation}
F_{\nu_{\mathrm{obs}}}(\nu^{\prime},t^{\prime})=F_{\nu_{\mathrm{obs}}}(\nu^{\prime} +
d\nu^{\prime},t^{\prime}-dt^{\prime})(\frac{\partial \dot{\gamma_{\mathrm{e}}^{\prime}} }
{\partial \gamma_{\mathrm{e}}^{\prime}} dt^{\prime}+1)+\frac{(1+z) \Gamma}{4 \pi D_L^2}\hat{Q}dt^{\prime}.
\label{eq:Pv1}
\end{equation}
Replacing $t^{\prime}$ with $t^{\prime}+dt^{\prime}$ and subtracting $F_{\nu_{\mathrm{obs}}}(\nu^{\prime},t^{\prime})$ from both sides of the equation, we can obtain
\begin{equation}
\begin{aligned}
F_{\nu_{\mathrm{obs}}}(\nu^{\prime},t^{\prime}+dt^{\prime})-F_{\nu_{\mathrm{obs}}}(\nu^{\prime},t^{\prime})
=& F_{\nu_{\mathrm{obs}}}(\nu^{\prime} +
d\nu^{\prime},t^{\prime})(\frac{\partial \dot{\gamma_{\mathrm{e}}^{\prime}}}
{\partial \gamma_{\mathrm{e}}^{\prime}} dt^{\prime}+1)\\
&-F_{\nu_{\mathrm{obs}}}(\nu^{\prime},t^{\prime})
+\frac{(1+z) \Gamma}{4 \pi D_L^2}\hat{Q}dt^{\prime}.
\label{eq:Pv2}
\end{aligned}
\end{equation}
Dividing both sides of Equation (\ref{eq:Pv2}) with $dt_{\mathrm{obs}}$, we have
\begin{equation}
\begin{aligned}
\frac{\partial F}{\partial t_{\mathrm{obs}}}
= \Lambda \frac{\partial F}{\partial \nu_{\mathrm{obs}}}+
\Omega F_{\nu_{\mathrm{obs}}}+\frac{\Gamma \mathcal{D}}{4\pi D_{\mathrm{L}}^{2}}\hat{Q},
\end{aligned}
\end{equation}
where
\begin{equation}
\Lambda=(\frac{3 q_{\mathrm{e}} B^{\prime}(1+z)\nu_{\mathrm{obs}}}{ \pi m_{\mathrm{e}}c\mathcal{D}})^{\frac{1}{2}}(\frac{\mathcal{D}}{1+z})^2\dot{\gamma_{\mathrm{e}}^{\prime}},
\end{equation}
\begin{equation}
\Omega=\frac{\mathcal{D}}{1+z}\frac{\partial \dot{\gamma_{\mathrm{e}}^{\prime}}}{\partial \gamma_{\mathrm{e}}^{\prime}},
\end{equation}
\begin{equation}
d t_{\mathrm{obs}}=(1+z) \Gamma(1-\beta_{\mathrm{j}}) d t^{\prime}.
\end{equation}
Noting that $\alpha = -\frac{\partial \ln F}{\partial \ln t_{\mathrm{obs}}}$ and $\beta= -\frac{\partial \ln F}{\partial \ln t_{\mathrm{obs}}}$, we will arrive at
\begin{equation}
\alpha= \frac{t_{\mathrm{obs}}}{\nu_{\mathrm{obs}}}\Lambda \beta-
\Omega t_{\mathrm{obs}}-\frac{t_{\mathrm{obs}}}{F_{\nu_{\mathrm{obs}}}}\frac{\Gamma \mathcal{D}}{4\pi D_{\mathrm{L}}^{2}}\hat{Q}.
\label{eq:A12}
\end{equation}

\bibliographystyle{aasjournal}
\bibliography{reference}

\begin{thebibliography}{}
\expandafter\ifx\csname natexlab\endcsname\relax\def\natexlab#1{#1}\fi
\providecommand{\url}[1]{\href{#1}{#1}}
\providecommand{\dodoi}[1]{doi:~\href{http://doi.org/#1}{\nolinkurl{#1}}}
\providecommand{\doeprint}[1]{\href{http://ascl.net/#1}{\nolinkurl{http://ascl.net/#1}}}
\providecommand{\doarXiv}[1]{\href{https://arxiv.org/abs/#1}{\nolinkurl{https://arxiv.org/abs/#1}}}

\bibitem[{{Barthelmy} {et~al.}(2005){Barthelmy}, {Cannizzo}, {Gehrels},
  {Cusumano}, {Mangano}, {O'Brien}, {Vaughan}, {Zhang}, {Burrows}, {Campana},
  {Chincarini}, {Goad}, {Kouveliotou}, {Kumar}, {M{\'e}sz{\'a}ros}, {Nousek},
  {Osborne}, {Panaitescu}, {Reeves}, {Sakamoto}, {Tagliaferri}, \&
  {Wijers}}]{Barthelmy05b}
{Barthelmy}, S.~D., {Cannizzo}, J.~K., {Gehrels}, N., {et~al.} 2005, \apjl,
  635, L133, \dodoi{10.1086/499432}

\bibitem[{{Blumenthal} \& {Gould}(1970)}]{Blumenthal70}
{Blumenthal}, G.~R., \& {Gould}, R.~J. 1970, Reviews of Modern Physics, 42,
  237, \dodoi{10.1103/RevModPhys.42.237}

\bibitem[{{Crumley} {et~al.}(2019){Crumley}, {Caprioli}, {Markoff}, \&
  {Spitkovsky}}]{Crumley19}
{Crumley}, P., {Caprioli}, D., {Markoff}, S., \& {Spitkovsky}, A. 2019, \mnras,
  485, 5105, \dodoi{10.1093/mnras/stz232}

\bibitem[{{Dai} \& {Lu}(1998)}]{Dai98}
{Dai}, Z.~G., \& {Lu}, T. 1998, \aap, 333, L87,
  \dodoi{10.48550/arXiv.astro-ph/9810402}

\bibitem[{{Deng} \& {Zhang}(2014)}]{Deng14}
{Deng}, W., \& {Zhang}, B. 2014, \apj, 785, 112,
  \dodoi{10.1088/0004-637X/785/2/112}

\bibitem[{{Eichler} {et~al.}(1989){Eichler}, {Livio}, {Piran}, \&
  {Schramm}}]{Eichler89}
{Eichler}, D., {Livio}, M., {Piran}, T., \& {Schramm}, D.~N. 1989, \nat, 340,
  126, \dodoi{10.1038/340126a0}

\bibitem[{{Eichler} \& {Waxman}(2005)}]{Eichler05}
{Eichler}, D., \& {Waxman}, E. 2005, \apj, 627, 861, \dodoi{10.1086/430596}

\bibitem[{Gao {et~al.}(2013)Gao, Lei, Zou, Wu, \& Zhang}]{Gao13a}
Gao, H., Lei, W.-H., Zou, Y.-C., Wu, X.-F., \& Zhang, B. 2013, New Astronomy
  Reviews, 57, 141, \dodoi{https://doi.org/10.1016/j.newar.2013.10.001}

\bibitem[{{Gao} {et~al.}(2021){Gao}, {Geng}, \& {Huang}}]{Gao21}
{Gao}, H.-X., {Geng}, J.-J., \& {Huang}, Y.-F. 2021, Astronomy \&
  Astrophysics,, 656, A134, \dodoi{10.1051/0004-6361/202141647}

\bibitem[{{Gao} {et~al.}(2022){Gao}, {Geng}, {Hu}, {Hu}, {Lan}, {Chang},
  {Zhang}, {Zhang}, {Huang}, \& {Wu}}]{Gao22}
{Gao}, H.-X., {Geng}, J.-J., {Hu}, L., {et~al.} 2022, \mnras, 516, 453,
  \dodoi{10.1093/mnras/stac2215}

\bibitem[{{Geng} {et~al.}(2018{\natexlab{a}}){Geng}, {Dai}, {Huang}, {Wu},
  {Li}, {Li}, \& {Meng}}]{Geng18c}
{Geng}, J.-J., {Dai}, Z.-G., {Huang}, Y.-F., {et~al.} 2018{\natexlab{a}},
  \apjl, 856, L33, \dodoi{10.3847/2041-8213/aab7f9}

\bibitem[{{Geng} {et~al.}(2018{\natexlab{b}}){Geng}, {Huang}, {Wu}, {Zhang}, \&
  {Zong}}]{Geng18b}
{Geng}, J.-J., {Huang}, Y.-F., {Wu}, X.-F., {Zhang}, B., \& {Zong}, H.-S.
  2018{\natexlab{b}}, \apjs, 234, 3, \dodoi{10.3847/1538-4365/aa9e84}

\bibitem[{{Geng} {et~al.}(2016){Geng}, {Wu}, {Huang}, {Li}, \& {Dai}}]{Geng16a}
{Geng}, J.~J., {Wu}, X.~F., {Huang}, Y.~F., {Li}, L., \& {Dai}, Z.~G. 2016,
  \apj, 825, 107, \dodoi{10.3847/0004-637X/825/2/107}

\bibitem[{{Geng} {et~al.}(2014){Geng}, {Wu}, {Li}, {Huang}, \& {Dai}}]{Geng14}
{Geng}, J.~J., {Wu}, X.~F., {Li}, L., {Huang}, Y.~F., \& {Dai}, Z.~G. 2014,
  \apj, 792, 31, \dodoi{10.1088/0004-637X/792/1/31}

\bibitem[{{Ghisellini} \& {Celotti}(1999)}]{Ghisellini99}
{Ghisellini}, G., \& {Celotti}, A. 1999, \apjl, 511, L93,
  \dodoi{10.1086/311845}

\bibitem[{{Giannios} \& {Spitkovsky}(2009)}]{Giannios09}
{Giannios}, D., \& {Spitkovsky}, A. 2009, \mnras, 400, 330,
  \dodoi{10.1111/j.1365-2966.2009.15454.x}

\bibitem[{{Gobat} {et~al.}(2023){Gobat}, {van der Horst}, \&
  {Fitzpatrick}}]{Gobat23}
{Gobat}, C., {van der Horst}, A.~J., \& {Fitzpatrick}, D. 2023, \mnras, 523,
  775, \dodoi{10.1093/mnras/stad1189}

\bibitem[{{Gould} \& {Schr{\'e}der}(1967)}]{Gould67}
{Gould}, R.~J., \& {Schr{\'e}der}, G.~P. 1967, Physical Review, 155, 1404,
  \dodoi{10.1103/PhysRev.155.1404}

\bibitem[{{Huang} {et~al.}(1999){Huang}, {Dai}, \& {Lu}}]{Huang99}
{Huang}, Y.~F., {Dai}, Z.~G., \& {Lu}, T. 1999, \mnras, 309, 513,
  \dodoi{10.1046/j.1365-8711.1999.02887.x}

\bibitem[{{Huang} {et~al.}(2000){Huang}, {Gou}, {Dai}, \& {Lu}}]{Huang00c}
{Huang}, Y.~F., {Gou}, L.~J., {Dai}, Z.~G., \& {Lu}, T. 2000, \apj, 543, 90,
  \dodoi{10.1086/317076}

\bibitem[{{Kobayashi} \& {Sari}(2000{\natexlab{a}})}]{Kobayashi00a}
{Kobayashi}, S., \& {Sari}, R. 2000{\natexlab{a}}, in American Institute of
  Physics Conference Series, Vol. 526, Gamma-ray Bursts, 5th Huntsville
  Symposium, ed. R.~M. {Kippen}, R.~S. {Mallozzi}, \& G.~J. {Fishman},
  550--554, \dodoi{10.1063/1.1361598}

\bibitem[{{Kobayashi} \& {Sari}(2000{\natexlab{b}})}]{Kobayashi00b}
{Kobayashi}, S., \& {Sari}, R. 2000{\natexlab{b}}, \apj, 542, 819,
  \dodoi{10.1086/317021}

\bibitem[{{Kumar} \& {Zhang}(2015)}]{Kumar15}
{Kumar}, P., \& {Zhang}, B. 2015, \physrep, 561, 1,
  \dodoi{10.1016/j.physrep.2014.09.008}

\bibitem[{{Li} {et~al.}(2019){Li}, {Geng}, {Huang}, \& {Li}}]{Li19}
{Li}, L.-B., {Geng}, J.-J., {Huang}, Y.-F., \& {Li}, B. 2019, \apj, 880, 39,
  \dodoi{10.3847/1538-4357/ab275d}

\bibitem[{{Longair}(2011)}]{Longair11}
{Longair}, M.~S. 2011, {High Energy Astrophysics}

\bibitem[{{Lundman} {et~al.}(2013){Lundman}, {Pe'er}, \& {Ryde}}]{Lundman13}
{Lundman}, C., {Pe'er}, A., \& {Ryde}, F. 2013, \mnras, 428, 2430,
  \dodoi{10.1093/mnras/sts219}

\bibitem[{{Margalit} \& {Quataert}(2021)}]{Margalit21}
{Margalit}, B., \& {Quataert}, E. 2021, \apjl, 923, L14,
  \dodoi{10.3847/2041-8213/ac3d97}

\bibitem[{{Martins} {et~al.}(2009){Martins}, {Fonseca}, {Silva}, \&
  {Mori}}]{Martins09}
{Martins}, S.~F., {Fonseca}, R.~A., {Silva}, L.~O., \& {Mori}, W.~B. 2009,
  \apjl, 695, L189, \dodoi{10.1088/0004-637X/695/2/L189}

\bibitem[{{Medina Covarrubias} {et~al.}(2023){Medina Covarrubias}, {De Colle},
  {Urrutia}, \& {Vargas}}]{Medina23}
{Medina Covarrubias}, R., {De Colle}, F., {Urrutia}, G., \& {Vargas}, F. 2023,
  arXiv e-prints, arXiv:2306.01136, \dodoi{10.48550/arXiv.2306.01136}

\bibitem[{{M{\'e}sz{\'a}ros} {et~al.}(1993){M{\'e}sz{\'a}ros}, {Laguna}, \&
  {Rees}}]{Meszaros93}
{M{\'e}sz{\'a}ros}, P., {Laguna}, P., \& {Rees}, M.~J. 1993, \apj, 415, 181,
  \dodoi{10.1086/173154}

\bibitem[{{M{\'e}sz{\'a}ros} \& {Rees}(1997)}]{Meszaros97}
{M{\'e}sz{\'a}ros}, P., \& {Rees}, M.~J. 1997, \apj, 476, 232,
  \dodoi{10.1086/303625}

\bibitem[{{M{\'e}sz{\'a}ros} \& {Rees}(2000)}]{Meszaros00}
---. 2000, \apj, 530, 292, \dodoi{10.1086/308371}

\bibitem[{{Paczy{\'n}ski}(1998)}]{Paczynski98}
{Paczy{\'n}ski}, B. 1998, \apjl, 494, L45, \dodoi{10.1086/311148}

\bibitem[{{Paczynski} \& {Rhoads}(1993)}]{Paczynski93}
{Paczynski}, B., \& {Rhoads}, J.~E. 1993, \apjl, 418, L5,
  \dodoi{10.1086/187102}

\bibitem[{{Paczynski} \& {Xu}(1994)}]{Paczynski94}
{Paczynski}, B., \& {Xu}, G. 1994, \apj, 427, 708, \dodoi{10.1086/174178}

\bibitem[{{Panaitescu}(2005)}]{Panaitescu05}
{Panaitescu}, A. 2005, \mnras, 363, 1409,
  \dodoi{10.1111/j.1365-2966.2005.09532.x}

\bibitem[{{Park} {et~al.}(2015){Park}, {Caprioli}, \& {Spitkovsky}}]{Park15}
{Park}, J., {Caprioli}, D., \& {Spitkovsky}, A. 2015, \prl, 114, 085003,
  \dodoi{10.1103/PhysRevLett.114.085003}

\bibitem[{{Pe'er}(2008)}]{Peer08}
{Pe'er}, A. 2008, \apj, 682, 463, \dodoi{10.1086/588136}

\bibitem[{{Pe'er} \& {Ryde}(2011)}]{Peer11}
{Pe'er}, A., \& {Ryde}, F. 2011, \apj, 732, 49,
  \dodoi{10.1088/0004-637X/732/1/49}

\bibitem[{{Pennanen} {et~al.}(2014){Pennanen}, {Vurm}, \&
  {Poutanen}}]{Pennanen14}
{Pennanen}, T., {Vurm}, I., \& {Poutanen}, J. 2014, \aap, 564, A77,
  \dodoi{10.1051/0004-6361/201322520}

\bibitem[{{Piran}(2004)}]{Piran04}
{Piran}, T. 2004, Reviews of Modern Physics, 76, 1143,
  \dodoi{10.1103/RevModPhys.76.1143}

\bibitem[{{Rees} \& {Meszaros}(1992)}]{Rees92}
{Rees}, M.~J., \& {Meszaros}, P. 1992, \mnras, 258, 41,
  \dodoi{10.1093/mnras/258.1.41P}

\bibitem[{{Rees} \& {M{\'e}sz{\'a}ros}(1994)}]{Rees94}
{Rees}, M.~J., \& {M{\'e}sz{\'a}ros}, P. 1994, \apjl, 430, L93,
  \dodoi{10.1086/187446}

\bibitem[{{Ressler} \& {Laskar}(2017)}]{Ressler17}
{Ressler}, S.~M., \& {Laskar}, T. 2017, \apj, 845, 150,
  \dodoi{10.3847/1538-4357/aa8268}

\bibitem[{{Rybicki} \& {Lightman}(1979)}]{Rybicki79}
{Rybicki}, G.~B., \& {Lightman}, A.~P. 1979, {Radiative processes in
  astrophysics}

\bibitem[{{Sari} \& {Piran}(1999)}]{Sari99}
{Sari}, R., \& {Piran}, T. 1999, \apj, 520, 641, \dodoi{10.1086/307508}

\bibitem[{{Sari} {et~al.}(1998){Sari}, {Piran}, \& {Narayan}}]{Sari98}
{Sari}, R., {Piran}, T., \& {Narayan}, R. 1998, \apjl, 497, L17,
  \dodoi{10.1086/311269}

\bibitem[{{Sironi} {et~al.}(2013){Sironi}, {Spitkovsky}, \& {Arons}}]{Sironi13}
{Sironi}, L., {Spitkovsky}, A., \& {Arons}, J. 2013, \apj, 771, 54,
  \dodoi{10.1088/0004-637X/771/1/54}

\bibitem[{{Spitkovsky}(2008{\natexlab{a}})}]{Spitkovsky08a}
{Spitkovsky}, A. 2008{\natexlab{a}}, \apjl, 673, L39, \dodoi{10.1086/527374}

\bibitem[{{Spitkovsky}(2008{\natexlab{b}})}]{Spitkovsky08b}
---. 2008{\natexlab{b}}, \apjl, 682, L5, \dodoi{10.1086/590248}

\bibitem[{{Spruit} {et~al.}(2001){Spruit}, {Daigne}, \& {Drenkhahn}}]{Spruit01}
{Spruit}, H.~C., {Daigne}, F., \& {Drenkhahn}, G. 2001, \aap, 369, 694,
  \dodoi{10.1051/0004-6361:20010131}

\bibitem[{{Thompson}(1994)}]{Thompson94}
{Thompson}, C. 1994, \mnras, 270, 480, \dodoi{10.1093/mnras/270.3.480}

\bibitem[{{Wang} {et~al.}(2015){Wang}, {Zhang}, {Liang}, {Gao}, {Li}, {Deng},
  {Qin}, {Tang}, {Kann}, {Ryde}, \& {Kumar}}]{Wang15b}
{Wang}, X.-G., {Zhang}, B., {Liang}, E.-W., {et~al.} 2015, \apjs, 219, 9,
  \dodoi{10.1088/0067-0049/219/1/9}

\bibitem[{{Warren} {et~al.}(2018){Warren}, {Barkov}, {Ito}, {Nagataki}, \&
  {Laskar}}]{Warren18}
{Warren}, D.~C., {Barkov}, M.~V., {Ito}, H., {Nagataki}, S., \& {Laskar}, T.
  2018, \mnras, 480, 4060, \dodoi{10.1093/mnras/sty2138}

\bibitem[{{Warren} {et~al.}(2017){Warren}, {Ellison}, {Barkov}, \&
  {Nagataki}}]{Warren17}
{Warren}, D.~C., {Ellison}, D.~C., {Barkov}, M.~V., \& {Nagataki}, S. 2017,
  \apj, 835, 248, \dodoi{10.3847/1538-4357/aa56c3}

\bibitem[{{Waxman}(1997)}]{Waxman97}
{Waxman}, E. 1997, \apjl, 491, L19, \dodoi{10.1086/311057}

\bibitem[{Wijers {et~al.}(1997)Wijers, Rees, \& Meszaros}]{Wijers97}
Wijers, R. A. M.~J., Rees, M.~J., \& Meszaros, P. 1997, Monthly Notices of the
  Royal Astronomical Society, 288, L51, \dodoi{10.1093/mnras/288.4.L51}

\bibitem[{{Yabe} {et~al.}(2001){Yabe}, {Xiao}, \& {Utsumi}}]{Yabe01}
{Yabe}, T., {Xiao}, F., \& {Utsumi}, T. 2001, Journal of Computational Physics,
  169, 556, \dodoi{10.1006/jcph.2000.6625}

\bibitem[{{Yu} \& {Dai}(2007)}]{YuDai07}
{Yu}, Y.~W., \& {Dai}, Z.~G. 2007, \aap, 470, 119,
  \dodoi{10.1051/0004-6361:20077053}

\bibitem[{{Zhang}(2018)}]{Zhang18}
{Zhang}, B. 2018, {The Physics of Gamma-Ray Bursts},
  \dodoi{10.1017/9781139226530}

\bibitem[{{Zhang} {et~al.}(2006){Zhang}, {Fan}, {Dyks}, {Kobayashi},
  {M{\'e}sz{\'a}ros}, {Burrows}, {Nousek}, \& {Gehrels}}]{Zhang06}
{Zhang}, B., {Fan}, Y.~Z., {Dyks}, J., {et~al.} 2006, \apj, 642, 354,
  \dodoi{10.1086/500723}

\bibitem[{{Zhang} \& {Yan}(2011)}]{ZhangYang11}
{Zhang}, B., \& {Yan}, H. 2011, \apj, 726, 90,
  \dodoi{10.1088/0004-637X/726/2/90}

\bibitem[{{Zhang} {et~al.}(2007){Zhang}, {Liang}, \& {Zhang}}]{Zhang07}
{Zhang}, B.-B., {Liang}, E.-W., \& {Zhang}, B. 2007, \apj, 666, 1002,
  \dodoi{10.1086/519548}

\bibitem[{{Zhang} {et~al.}(2019){Zhang}, {Geng}, \& {Huang}}]{Zhang19}
{Zhang}, Y., {Geng}, J.-J., \& {Huang}, Y.-F. 2019, \apj, 877, 89,
  \dodoi{10.3847/1538-4357/ab1b10}

\end{thebibliography}

\end{document}